\documentclass[
reprint,
nofootinbib,
 amsmath,amssymb,
 aps,
]{revtex4-2}
\usepackage[dvipsnames,table,xcdraw]{xcolor}
\usepackage{graphicx}
\usepackage{bm}
\usepackage{tikz}
\usepackage{braket}
\usepackage{dsfont}
\usepackage{hyperref}
\hypersetup{
    colorlinks=true, 
    linktoc=all,     
    linkcolor=blue,  
    citecolor=magenta
}
\usepackage{bbold}
\usepackage{soul}

\usetikzlibrary{decorations.pathreplacing,calligraphy}
\usetikzlibrary{decorations.markings}

\newcommand{\prare}{p_{\rm rare}}
\newcommand{\pfail}{\mathbb{P}_{\rm fail}}
\newcommand{\psucc}{\mathbb{P}_{\rm succ}}

\newcommand{\nn}{\nonumber \\}

{}
\newcommand{\mycomment}[1]{}

\makeatletter
\def\l@subsection#1#2{}
\def\l@subsubsection#1#2{}
\makeatother

\begin{document}

\preprint{APS/123-QED}

\title{Rare Events and Griffiths Phases in Topological Quantum Error Correction}

\author{Adithya Sriram, Nicholas O'Dea, Yaodong Li, Tibor Rakovszky, Vedika Khemani}
\affiliation{Department of Physics, Stanford University, Stanford, CA 94305}
\date{March 17, 2025}

\begin{abstract}

The performance of quantum error correcting (QEC) codes are often studied under the assumption of spatio-temporally uniform error rates. On the other hand, experimental implementations almost always produce heterogeneous error rates, in either space or time, as a result of effects such as imperfect fabrication and/or cosmic rays.
It is therefore important to understand if and how their presence can affect the performance of QEC in qualitative ways.
In this work, we study effects of non-uniform error rates in the representative examples of the 1D repetition code and the 2D toric code, focusing on when they have extended spatio-temporal correlations; these may arise, for instance, from rare events (such as cosmic rays) that temporarily elevate error rates over the entire code patch.
These effects can be described in the corresponding statistical mechanics models for decoding, where long-range correlations in the error rates lead to \textit{extended rare regions} of weaker coupling.
For the 1D repetition code where the rare regions are linear,  
we find two distinct decodable phases: a conventional \textit{ordered phase} in which logical failure rates decay exponentially with the code distance, and a rare-region dominated \textit{Griffiths phase} in which failure rates are parametrically larger and decay as a stretched exponential. 
In particular, the latter phase is present when the error rates in the rare regions are \textit{above} the bulk threshold.
For the 2D toric code where the rare regions are planar, we find no decodable Griffiths phase: rare events which boost error rates above the bulk threshold lead to an asymptotic loss of threshold and failure to decode. 
Unpacking the failure mechanism implies that techniques for suppressing \textit{extended} sequences of repeated rare events (which, without intervention, will be statistically present with high probability) will be crucial for QEC with the toric code.

\end{abstract}

\maketitle
\tableofcontents

 \section{Introduction}

A quantum error correcting (QEC) code aims to protect quantum information against the effects of environmental decoherence by robustly and redundantly encoding logical qubits into entangled states of many physical qubits.
The threshold theorem of fault-tolerant quantum computing allows for arbitrarily long computations to be performed on the logical qubits with arbitrarily high accuracy; i.e., the \textit{logical error rate} goes to zero as the number of physical qubits is scaled up, provided that noise rates are below the \textit{error threshold}, whose value depends on both the code being used and the types of errors that occur~\cite{aharonov1999FT, gottesman1999FT}.

Accurately estimating the threshold and the logical failure rates is of great importance for any hardware implementation.
While this often requires taking into consideration various details (including a detailed description of the error model), it is believed that much of the qualitative aspects of QEC can be understood within simplified toy models, involving only \textit{phenomenological noise}.
Within this setting, mappings to statistical mechanics models of decoding~\cite{Dennis_2002,Chubb_2021,bombin2013introduction, PhysRevX.9.021041} have been particularly useful, in informing both the theoretical upper bound on the threshold, and the asymptotic scaling of the sub-threshold logical failure rate.

Phenomenological noise is often studied under the simplifying assumption of uniform error rates on all qubits for all times, while realistic implementations almost always display some sort of heterogeneity, resulting in e.g. non-uniform noise rates \cite{Tiurev2023correctingnon}.
It is conceivable that when the fluctuations are short-range correlated, the simplifying assumption is still valid.
On the other hand, with extended spatio-temporal correlations --- which naturally occur due to effects ranging from fabrication errors~\cite{PhysRevA.106.062428} to stochastic events such as cosmic rays striking the quantum hardware \cite{McEwen_2021, harrington2024synchronous, mcewen2024resisting} --- 
it is unclear if and how things will be qualitatively different. Furthermore, error bursts with extended spatio-temporal correlations can occur due to sources not yet identified. For instance, in a recent demonstration of their platform's new capabilities, experimentalists at Google identified a logical noise floor due to error bursts, but could not identify what the source of these error bursts were \cite{acharya2024quantum}.
Studies of such effects are therefore needed for a qualitative understanding of code performance in practice~\cite{Stace_2009, Stace_2010, Auger_2017, Chubb_2021, Strikis_2023, aasen2023faulttolerant, Tiurev2023correctingnon, Huo_2017, Nickerson_2019}. 

In this work, we investigate said effects of non-uniform spatio-temporally correlated noise rates 
which map to \textit{extended, sub-dimensional rare regions} in the corresponding stat mech models.
We show that such rare regions, despite being sub-dimensional and rare, can often have a disproportionately large effect and dominate the logical failure of the code, a phenomenon known as a \textit{Griffiths effect}.
Griffiths effects have been extensively studied in disordered systems, and correspond to rare but large disorder fluctuations which dramatically change the properties of phase transitions (and the proximate phases)~\cite{Vojta_2006}.

We mainly focus on rare events (such as cosmic rays) that temporally increase the error rate from $p_{\rm bulk}$ to $p_{\rm rare} > p_{\rm bulk}$ over the entire code patch.
We study this for the representative examples of the 1D repetition code and the 2D toric code, in an phenomenological error model with both measurement and qubit errors.
They are described by the 2D random bond Ising model (RBIM) and the 3D random plaquette Ising gauge theory (RPGT), respectively, both with non-uniform coupling strengths, see Sec.~\ref{sec:background}. {Fabrication error correspond to elevated error rates in a timelike quasi-$1d$ region in spacetime, rather than an elevated error rate over the entire code path at a given time. }

In the case of the the 1D repetition code, we leverage known results from disorder physics to argue that the rare regions lead to a new \textit{Griffiths phase} above the conventional decodable phase, where the logical failure rate decays as a stretched exponential of the code distance, instead of the exponential scaling that occurs in the conventional decodable phase (when rare regions are absent).
Therefore, while the rare regions do not lead to a failure of error correction (there is still a finite threshold), they can qualitatively change the code's performance by parametrically increasing the logical failure rate.  
Our results are described in detail in Sec.~\ref{sec:linear}.

In contrast, for the 2D toric code, rare events can be catastrophic: we find that as soon as $p_{\rm rare}$ is above the bulk threshold, the entire code \textit{loses its threshold} (i.e. the probability of a logical error no longer vanishes in the thermodynamic limit)\footnote{We often have in mind a situtation where $p_{\rm bulk}$ can be tuned by improving the qubit quality, whereas $p_{\rm rare}$ is an external parameter that we have no control over.
Therefore, we say the effects are catastrophic if there is a large enough $\prare$ which makes the code undecodable even when $p_{\rm bulk} \to 0$.
However, it might be possible to gain some control over $p_{\rm rare}$ in the hardware, as pointed out in Ref.~\cite{mcewen2024resisting}.}. Notably, this happens independently of the bulk error rate, i.e. even as $p_{\rm bulk} \rightarrow 0$. We note that similar error models were recently studied numerically by Tan et.~al.~\cite{tan2024resilience}, where it was assumed that  rare events such as cosmic rays last for at most a \textit{finite duration}. Within their setup, the rare events are found to be ``benign'', in the sense that when the rare region error rate $p_{\rm rare}$ is slightly above the bulk threshold, the code can remain in the decodable phase by correspondingly decreasing the bulk error rate $p_{\rm bulk}$. {This is also similar to another model considered in ~\cite{ramette2023faulttolerantconnectionerrorcorrectedqubits}, where the authors considered noisy toric code patches linked together by noisy "seams" in order to communicate logical information between code patches. There also, as the seams are always of lower dimension than the bulk, the system could tolerate high noise rates on the seams compared to the bulk.} The key difference in our work is that we consider a more \textit{realistic} setup where the rare events occur at a  \textit{finite rate} in time. This, in turn, implies that the longest period with elevated error rates is \textit{typically unbounded from above}: an $L\times L$ toric code system requires $O(L)$ rounds of repeated measurements for decoding, which typically produces a largest rare-event sequence of size $O(\log(L))$.  We argue that these largest rare regions dominate the logical failure rate, and lead to loss of threshold. Both our setup and that of Ref.~\cite{tan2024resilience}  can be understood within our theoretical framework, as we detail in Sec.~\ref{sec:planar}.
Our results imply that techniques for suppressing lasting rare events~\cite{mcewen2024resisting} will be crucial to QEC with the toric code.

We also provide a physical interpretation of the difference between the 1D repetition code and the 2D toric code as follows.
Both models have excitations that are pointlike, and logical errors that are one-dimensional.
It is convenient to consider their dual models under Kramers-Wannier duality, after the random sign disorder in the stat mech models are neglected.
We obtain 2D and 3D Ising models with non-uniform couplings, respectively, where logical failure of the code can be related to dual Ising correlation functions in both cases.
The difference between the two phase diagrams can be attributed to whether the rare regions can order by themselves in isolation, in this dual picture. 

{
In our work, we consider only a time-varying bit-flip error rate for simplicity; all our results carry over to the case where the measurement error rate also experiences error bursts. Furthermore, we expect that the physics that we explore in this phenomenological noise model carries over to the case of circuit level noise as well. That is, if the gates used to extract the syndrome also are affected due to the error burst, this is well captured by an elevated phenomenological measurement error rate.
}

Finally, we note that our results for the 1D repetition code are obtained by leaning on known results for Griffiths physics in the celebrated McCoy-Wu model, which is a disordered 2D Ising model with correlated columnar disorder~\cite{PhysRev.176.631}. In contrast, the toric code problem yields a 3D RPGT with correlated disorder, which has not been studied much before in the context of Griffiths physics. Thus our analysis of this problem should also be of independent interest as a stat mech problem.

\begin{figure}
    \centering
    \includegraphics[width=\linewidth]{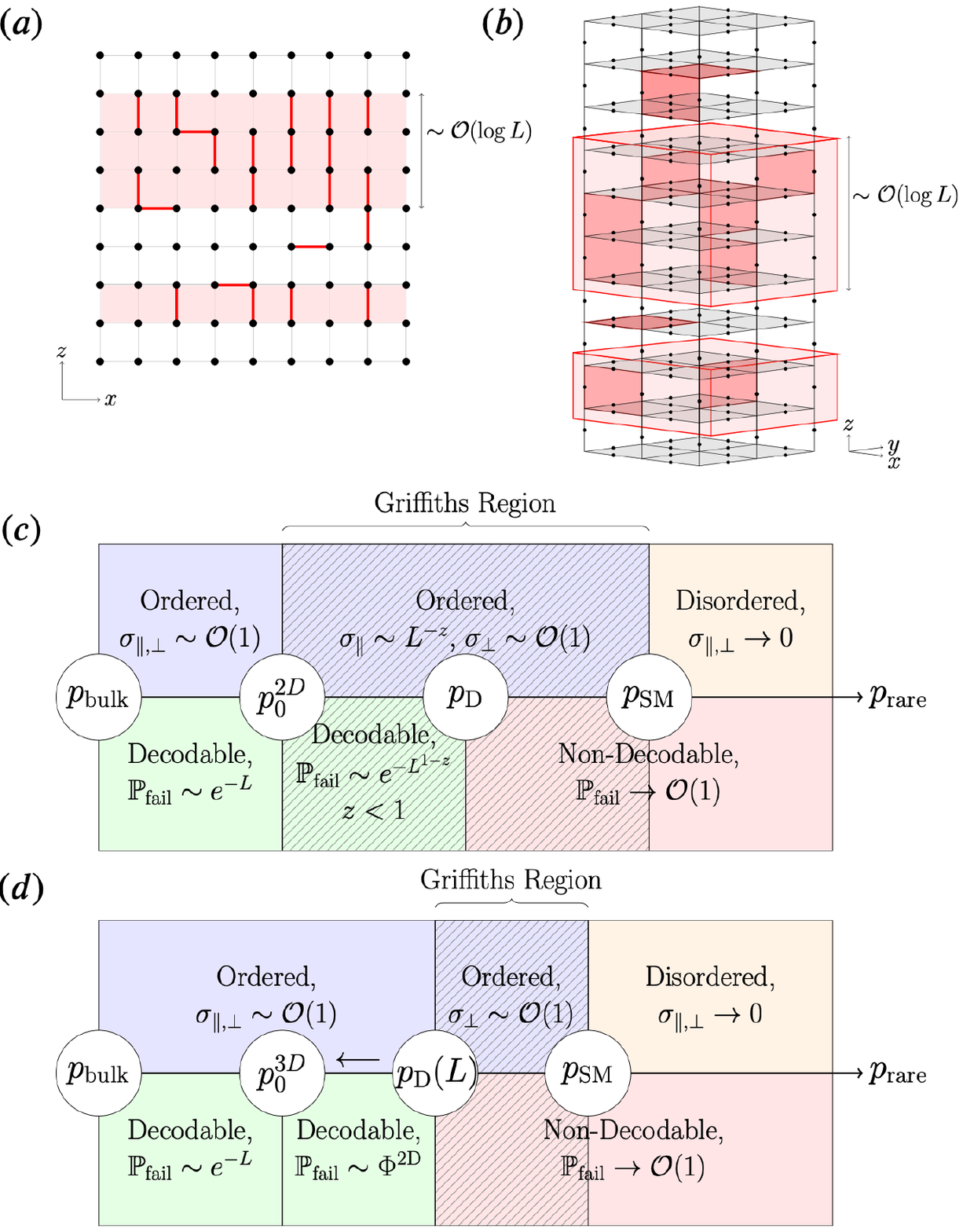}
    \caption{ Statistical mechanics models:  (a) a 2D RBIM corresponding to the decoding problem of the repetition code and (b) a 3D RPGT corrsponding to the toric code.  The horizontal $x/y$ directions are the spatial directions of the codes, and the vertical $z$ direction is time. Both models have time varying bit flip errors. The red shaded regions are rare regions where the error rate is higher, i.e. where the vertical bond/plaquette coupling strength is weaker. Bit flip and measurement errors are shown via thick red links/dark shaded plaquettes. A higher density of them is located in the rare regions. The largest rare region is of height $\sim \mathcal{O}(\log L)$. (c) and (d) are the stat mech and decoding phase diagrams of the models in (a) and (b) respectively as a function of $p_{\rm rare}$ (see main text). 
    }
    \label{fig:cartoons}
\end{figure}

\section{Background and Models\label{sec:background}}


We will focus on two familiar examples of error correcting codes, the 1D repetition code and the 2D toric / surface code. Both are characterized by a set of stabilizer parity checks. Errors are characterized by their syndrome, the set of checks they violate. In both examples, these are point-like excitations (domain walls and anyons, respectively). The goal of error correction is to pair up these excitations in a way that undoes the effect of the error. One also has to deal with errors in the measurement of the stabilizers, which is usually done by combining information from many rounds of measurements; this introduces a time coordinate, making the two problems 1+1 and 2+1 dimensional, respectively. 

For any given error correcting algorithm, the \textit{logical failure rate} refers to the probability that the algorithm fails to correctly recover the encoded logical information. It is related to the probability of large error chains whose syndromes are incorrectly paired up by the algorithm. The \textit{error threshold} of the code is the maximal error rate (of both physical and measurement errors) such that the failure rate goes to zero as the number of (qu)bits is taken to infinity. Of particular interest is the \textit{maximum likelihood} decoder, which corresponds to the theoretically optimal decoding and thus yields the largest threshold. 

While usually both the physical and the measurement error rates are taken to be constants, which are the same everywhere in the system, and for all the different measurement rounds, in realistic scenarios, they would be different for different qubits / stabilizers and can also fluctuate in time. We will be interested in the effect of such fluctuations, particularly those with long-range correlations in space or time (focusing on the former case).

\subsection{Statistical mechanical models}\label{sec:statmechmodels}

The threshold and failure rate of stabilizer codes can be understood in terms of appropriate disordered statistical mechanics models~\cite{Dennis_2002,Chubb_2021,bombin2013introduction, PhysRevX.9.021041}. 
Detailed derivations of these models are of secondary importance for our purposes, and we refer the reader to \cite{Dennis_2002} for a more thorough explanation.
Throughout this paper, we focus on the illustrative examples of the 1D repetition code and the 2D toric code.
As both are CSS codes, we look solely at the $Z$ stabilizers of the stabilizer code with stochastic single-qubit $X$ noise acting on the system.

For the 1D repetition code, the stat mech model associated with decoding is known to be a 2D random bond Ising model (RBIM) on a square lattice,
\begin{align}\label{eq:rbim}
    Z_{\rm RBIM} = \sum_{\{ \sigma \}} \exp \left(\sum_{\langle j,k \rangle} K_{jk} \tau_{jk} \sigma_j \sigma_k \right),
\end{align}
and for the 2D toric code, the model is a 3D random plaquette (Ising) gauge theory (RPGT),
\begin{align}\label{eq:rpgt}
    Z_{\rm RPGT} = \sum_{\{ \sigma \}} \exp \left(\sum_\Box K_\Box \tau_\Box \prod_{\langle j, k\rangle \in \Box} \sigma_{jk} \right).
\end{align}
In Eq.~\eqref{eq:rbim}, the sum is over edges $jk$ containing nearest-neighbor spins $\sigma_j, \sigma_k$.
In Eq.~\eqref{eq:rpgt}, the sum is over plaquettes and each term is a product over spins which reside on the edges $jk$ of the plaquette.
In both models, there is one Ising spin per constant time plane for each physical qubit.
Couplings that extend into the time direction (timelike) correspond to bitflip errors, and in Fig.~\ref{fig:cartoons} they are represented as either vertical bonds (for the repetition code, see Fig.~\ref{fig:cartoons}(a)) or plaquettes with normals pointing in spatial directions (for the toric code, see Fig.~\ref{fig:cartoons}(b)).
Couplings within constant time planes (spacelike) take the form of $Z$ stabilizers and correspond to measurement errors.
In Fig.~\ref{fig:cartoons} they are represented by horizontal bonds (for the repetition code, see Fig.~\ref{fig:cartoons}(a)) or plaquettes with normals pointing in the time $z$ direction (for the toric code, see Fig.~\ref{fig:cartoons}(b)).

For the ease of discussion, in the following we use $\alpha$ as a label for couplings, to refer to either a bond $\langle j, k \rangle$ in the RBIM, or a plaquette $\Box$ in the RPGT. We denote the local error rate by $p_\alpha$ 
(i.e. it is a measurement error rate for a spacelike coupling, and a physical error rate for a timelike coupling). The local coupling strengths $K_\alpha$ are related to the local error rates via the \textit{Nishimori condition}
\begin{align} \label{eq:nishimori}
    e^{-2 K_\alpha} = \frac{p_\alpha}{1-p_\alpha}.
\end{align}
Here, this equation is understood to hold for each and every $\alpha$, in both Eqs.~(\ref{eq:rbim},\ref{eq:rpgt}). 
The $\tau_\alpha$ variables incorporate the error history and introduce quenched random sign disorder into the system.
They take the value $\tau_\alpha=-1$ when an error happens on $\alpha$ i.e. with probability $p_\alpha$, and take the value $\tau_\alpha=+1$ with probability $1-p_\alpha$.


As we are interested in error models with spatio-temporal heterogeneity, we introduce a second type of randomness, and allow the local error rates $p_\alpha$ to differ for different $\alpha$, while maintaining the Nishimori condition Eq.~\eqref{eq:nishimori} everywhere.
Correspondingly, the coupling constants $K_\alpha$ also vary in spacetime. Thus, we will study stat mech models with two types of disorder, namely (i) random sign disorder $\tau_\alpha$, and (ii) spatio-temporally varying $K_\alpha$.
Physically, (i) can be understood as coming from different stochastic error realizations within a \textit{fixed} error model, whereas (ii) is due to randomness in the error model itself, defined by the error rates at all locations $\{p_\alpha\}$.

\subsection{Defects and failure rates}

\begin{figure}
    \centering
    \includegraphics[width=\linewidth]{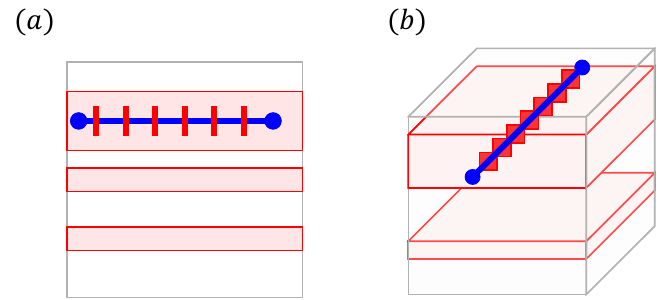}
    \caption{The relevant defects for the models shown in Fig.~\ref{fig:cartoons}(a,b). Panel (a) depicts the repetition code where the defect is a line of flipped vertical bonds corresponding to a domain wall. Panel (b) depicts the toric code where defect is a line of flipped plaquettes, also called a flux tube. Both defects are shown in rare regions where they are more likely to occur. Logical errors correspond to large defects, of sizes comparable to the linear system size $L$}
    \label{fig:defects}
\end{figure}

Condition~\eqref{eq:nishimori} ensures that the stat mech models correctly encode the success of maximal-likelihood decoders that have full knowledge of the error model.
In particular~\cite{Dennis_2002}, the failure and success probabilities of decoding are related to the free energy cost of the topological defect in the stat mech model which corresponds to a logical error.
For the RBIM, this is a domain wall; and for the RPGT, this is a magnetic flux tube (see Fig.~\ref{fig:defects}), both extending perpendicular to the temporal direction.
As the defects in both models are linelike, they have a free energy cost which to leading order scales as 
\begin{align} \label{eq:Delta_F_def}
    \Delta F \propto \sigma(\ell) \cdot \ell, 
\end{align}
where $\ell$ is the length of the defect and $\sigma(\ell)$ is its tension. In the ordered phases of the stat mech model (ferromagnetic for RBIM, deconfined for RPGT), defects are costly with diverging free-energy cost. This maps to the decodable phase with success probability approaching 1 as the system size is scaled up. A given defect may or may not correspond to a logical error. For instance, in the RBIM mapping of the 1+1D repetition code, only the defect in the horizontal direction is a logical operator (see Fig. ~\ref{fig:defects}).\footnote{Throughout this work, we denote the line tension of defects associated with logical operators as $\sigma_\parallel$, as they are parallel to the rare regions. We denote the line tension associated with the \textit{transversal} defect going in the temporal direction as $\sigma_\perp$, for it is perpendicular to the rare regions.}
In codes with uncorrelated disorder, these quantities are expected to scale the same in all phases. However, as we will see below, this need not be the case for long-range correlated disorder.

For a given error model $\{p_\alpha \}$, which thereby fixes the set $\{ K_\alpha \}$, the relative success probability for this error model is given by~\cite{Dennis_2002, kovalev2013spin, li2024perturbative}
\begin{align}\label{eq:defectcost}
    \pfail(\{K_\alpha\}) = 
    \left[  \frac{e^{- \Delta F}}{1+e^{- \Delta F}}  \right]_{\{\tau_\alpha\}},
\end{align}
where $\Delta F$ is the free energy cost of the defect within a particular realization of $\{ \tau_\alpha \}$ and $\left[ \ldots \right]_{\{\tau_\alpha\}}$ denotes the quenched average over $\{\tau_\alpha\}$.




Numerically, we use the minimium weight perfect matching (MWPM) decoder as implemented through the PyMatching package~\cite{higgott2023sparse, supp} to calculate $\pfail$.
Though many of our physical arguments are derived from considerations of the maximum likelihood decoder, in App.~\ref{sec:app_MWPM_griffiths} we provide arguments for why they are also expected to hold for the MWPM decoder.

We calculate $\pfail$ for each error model $\{ p_{\alpha} \}$.
This process yields a distribution of $\pfail$.
We do not expect $\pfail$ to be self-averaging, as it will be dominated by rare error models where the LRR are exceptionally large, see App.~\ref{sec:app_MWPM_griffiths} for further discussions.
Instead of the mean $\pfail$, we will mostly use the median $\pfail$ in our numerics.
From this we may \textit{numerically} define the defect cost $\Delta F$ as
\begin{align}\label{eq:deltaFdef}
    \Delta F \equiv -\ln \left[ \frac{(\pfail)_{\rm med}}{1- (\pfail)_{\rm med} }\right],
\end{align}
in analogy with Eq.~\eqref{eq:defectcost}.\footnote{Given the average over the sign disorder $\{\tau_a\}$ in Eq.~\eqref{eq:defectcost}, $\Delta F$ in Eq.~\eqref{eq:deltaFdef} is not the median over error rate realizations of the sign-disorder-averaged free energy. This does not matter for our purposes.}

\subsection{Rare regions \label{sec:rareregioneffects}} 

Our discussions so far have been fairly general.
Now, we turn to our primary focus, that is \textit{long range} correlations in the distribution of non-uniform error rates, and the effect of these on the free-energy cost of defects. 
For concreteness, we focus on cases with uniform measurement error rates $p_{\rm meas}$.
They correspond to uniform spacelike couplings in the stat mech model.
In contrast, we set bit-flip error rates $p_{\rm bf}$ to be spatially uniform on each time slice, but temporally varying between time slices, and are drawn from a Bernoulli distribution:
\begin{align}\label{eq:bernoulli}
    p_{\rm bf}(t) = \begin{cases} 
      p_{\text{bulk}} & \text{with probability }  1 - \gamma, \\
      p_{\text{rare}} & \text{with probability }  \gamma .
   \end{cases}
\end{align} 
These produce varying timelike couplings in the stat mech model via Eq.~\eqref{eq:nishimori}, see Fig.~\ref{fig:cartoons}.
This is a minimal model of stochastic rare events  which globally affect the system, inspired by phenomena such as cosmic rays. 

Let us denote by $p_0$ the threshold of the code when the error rate is uniform, i.e. when $\prare = p_{\rm bulk}$.
(This number is set by $p_{\rm meas}$, but we omit this dependence here.) In App.~\ref{sec:clean}, we calculate $p_0$ for the models use in this work.
We are interested in the following regime,
\begin{align} \label{eq:pbulk_prare_relative_to_p0}
    p_{\rm bulk} < p_0 < \prare,
\end{align}
which models an experiment that would be below threshold if not for rare events that temporarily elevates the error rate to $\prare$.
Parts of the stat mech model where $p_{\rm bf} = \prare$ are henceforth referred to as \textit{rare regions}.

Whenever $\gamma > 0$, 
we have rare regions whose \textit{typical} temporal extent $L_\perp$ is finite, which should be benign from an error correction point of view.
However, 
we argue below that logical failure events will be dominated by the \textit{largest rare region} (LRR), and with $\gamma > 0$ the largest $L_\perp$ will typically be unbounded from above in an infinitely long experiment.\footnote{\label{fn:temporal_extent}Throughout the paper, we will take the time for the experiment (hence also the ``height'' of the stat mech models) to be proportional to $L$.}
In Sec.~\ref{sec:linear} and Sec.~\ref{sec:planar} below, we present this rare region analysis for the 1D repetition code and for the 2D toric code, respectively, using both analytic arguments and numerics.

\section{Linear Rare Regions for the 1+1D Repetition Code \label{sec:linear}}



In this section, we analytically predict the effects of the correlated disorder in bitflip rates Eq.~\eqref{eq:bernoulli} in the 1D repetition code using the model described in Sec.~\ref{sec:statmechmodels}, and we numerically test these predictions.
The repetition code is a classical code that can serve as a simple theoretical model and as a experimental benchmark~\cite{google2021exponential, acharya2024quantum}.
In particular, here, the repetition code allows an analogy with the McCoy-Wu model, where the asymptotic scaling of the defect free energy can be predicted analytically \cite{PhysRev.176.631, PhysRevB.51.6411}.

As we noted in the previous section, the stat mech model describing decoding the 1D repetition code of length $L$ is an $L \times T$ RBIM for a time duration of $T$; we take $T=L$ in the following.
Logical failure rates are controlled by defect free energy costs, and the defects in the RBIM are domain walls.
The correlated disorder in bitflip rates introduces rows of weak vertical bonds in the RBIM that also have a higher likelihood of sign errors; we call regions of consecutive weak bonds in time \textit{rare regions}.



\subsection{Central assumption about largest rare region \label{sec:assumption}}

A central simplifying assumption we make in order to derive a qualitative phase diagram is that the logical failure of the code is dominated by the largest rare region (LRR) of the model\footnote{By ``dominate," we do not require the LRR to strictly set the logical failure rates; rather, we only assume that studying the LRR will give correct predictions for the asymptotic behavior in the phases on either side of the threshold, even if the LRR by itself does not necessarily predict nonuniversal quantities like phase boundaries in the full $L \times L$ model.}.
Correspondingly, in the stat mech model, we assume that the cost of the spatial defect $\Delta F$ is controlled by the LRR of weak couplings.
We will see that much of the numerical results can be understood qualitatively by focusing on the LRR and treating it as an isolated system. 

Our assumption can be understood from the intuitive picture that logical errors will most likely occur when the error rates are above threshold for the longest duration. One might also anticipate the assumption directly from the stat mech model by noting that the LRR has the most ``room'' for the domain wall to move (relative to smaller rare regions), and therefore the defect gains the most entropy (note that, since the bulk couplings are larger, it is energetically favorable for the defect to stay within rare regions). Both of these pictures neglect the interaction between the LRR and the bulk, as well as between LRR and other rare regions. We will find that much of the phase diagram is qualitatively described by LRR, although we note that there are discrepancies at large $\prare$ sufficiently far above threshold.

We briefly summarize the leading order scaling of defect free energy in \text{clean} models where neither the random sign disorder nor the spatio-temporal randomness in $\{K_\alpha\}$ are present.
These clean models are a convenient toy picture for the LRR, and provide a basis for comparison with our numerical results of $\Delta F$.


For the LRR in the 1D repetition code, we obtain the 2D Ising model living in a thin strip of dimensions $L \times L_\perp$.
In the paramagnetic phase with $K_{\rm rare} < K_c$, a high-temperature expansion can be used to show that the line tension decays to zero with increasing $L_\perp$ as
\begin{align} \label{eq:sigma_exp_L_perp}
    \sigma_\parallel(L) \propto e^{-\alpha L_\perp},
\end{align}
where $\alpha$ depends continuously on $K_{\rm rare}$. We discuss this scaling in more detail in Appendices~\ref{sec:app_2dHTE} and~\ref{sec:app_2dKW}.

For a rate of weak couplings $\gamma > 0$ and for a $O(L)$ height of the stat mech model, we have that the typical $L_\perp$ for the LRR grows as
\begin{align} \label{eq:LRR_height}
    L_\perp \propto \log L,
\end{align}
with a nonzero constant of proportionality. $L_\perp$ is distributed identically to the longest run of heads in a sequence of $L$ biased coin flips, which is known to have mean $\propto \log(L)$ up to $O(1)$ corrections~\cite{gordon1986extreme}. This random variable also has just $O(1)$ variance, meaning that its mean and median behave similarly. The mean $\propto \log(L)$ can be understood heuristically as striking a balance between $L$ opportunities to have a length $L_\perp$ string of rare regions and the exponentially decaying probability of such a consecutive string: the expected number of rare regions of height $L_\perp$ is asymptotically proportional to $L \gamma^{L_\perp}$, which is one for $L_\perp \sim \log L$.

Together, Eqs.~\ref{eq:sigma_exp_L_perp} and \eqref{eq:LRR_height} reproduce the same scaling as in Eq.~\eqref{eq:sigma_L_powerlaw_z}. The value of $z$ depends on several factors. It depends on the height of the rare region (and hence $\gamma$, which sets the proportionality constant in Eq.~\eqref{eq:LRR_height}). It also depends on how deep into the paramagnetic phase the rare region is (and hence on $K_\text{rare}$ and hence $p_\text{rare}$, which sets $\alpha$ in Eq.~\eqref{eq:sigma_L_powerlaw_z}). 

\subsection{Comparison to McCoy-Wu}
A useful comparison can be made between the stat mech model for the 1D repetition code and the McCoy-Wu model~\cite{PhysRev.176.631}.
The latter can be obtained from Eq.~\eqref{eq:rbim} by removing the sign disorder (setting $\tau_{jk} = +1$ everywhere) but still keeping the correlated disorder in the couplings $\{K_\alpha\}$.
By analogy with Eq.~\eqref{eq:pbulk_prare_relative_to_p0}, we are interested in the regime, 
\begin{align}
    K_{\rm bulk} > K_0^{2D} > K_{\rm rare},
\end{align}
where $K_0^{2D}$ is the critical coupling strength of the 2D uniform Ising model without any weak bonds.
The rare regions of the model are locally in the paramagnetic phase, whereas the bulk is still in the ferromagnetic phase, and the McCoy-Wu model is said to be in the \textit{Griffiths phase}.
Refs.~\cite{PhysRevLett.69.534, PhysRevB.51.6411} showed that a defect parallel to the rare region direction has a line tension of the form (compare Eq.~\eqref{eq:Delta_F_def})
\begin{align} \label{eq:sigma_L_powerlaw_z}
    \sigma_\parallel(L) \propto L^{-z},
\end{align}
where $L$ is the linear size of the system and $z$ is a dynamical exponent which depends continuously on $K_{\rm rare}$~\cite{PhysRevB.51.6411}. $z=0$ when $K_\text{rare} = K_0^{2D}$. 

Unlike the LRR approximation, $z$ tends to infinity at a finite value of $K_\text{rare} = K_c^{MW}$. This point coincides with the loss of spontaneous magnetization in the McCoy-Wu model, and is typically identified as the critical point of the McCoy-Wu model. By spontaneous magnetization, we mean the magnetization in the thermodynamic limit on taking a bulk longitudinal field $h \to 0^+$. Note that other common metrics of order may show transitions that do not necessarily coincide with $K_c^{MW}$; vertical correlation functions that cut perpendicular to the rare regions can still decay to zero for some choices of $K>K_c^{MW}$. 

For $K_\text{rare} < K_c^{MW}$, the domain wall tension $\sigma_\parallel$ decays faster than any power law. This is not captured in the LRR approximation. 


\subsection{Predicted phase diagram}
Using intuition from the McCoy-Wu model, we can now piece together the phase diagram shown in  Fig.~\ref{fig:cartoons}(c) as a function of $\prare$ at fixed $p_{\rm bulk} < p_0^{2D}$. Here, $p_0^{2D}$ is the critical bitflip error rate of the $L \times L$ repetition code. This should be distinguished from the threshold $p_\text{D}$ and the critical $p_\text{SM}$.

When we write $\pfail$ in this section, we consider the median $\pfail$.

First, when $p_{\text{rare}} < p_0^{2D}$, the entire system is in the decodable phase and the code is necessarily decodable. The RBIM exhibits ordinary ferromagnetic behavior such as a finite domain wall tension. This corresponds to a failure rate $\pfail$ decaying exponentially in system size.

For $p_{\text{rare}}$ between $p_0^{2D}$ and $p_{\rm SM}$, the rare regions are in the paramagnetic phase, but the stat mech model remains ordered. The stat mech model is in the ferromagnetic Griffiths phase with $\sigma_\parallel$ described by Eq.~\eqref{eq:sigma_L_powerlaw_z}, giving a domain wall cost of the form $\Delta F \approx L \sigma_\parallel(L) \propto L^{1-z}$. 
We expect $z$ to vary continuously with $p_{\rm rare}$. In particular, $z=0$ when $\prare = p_0^{2D}$. This phase is denoted by the hatched region in Fig. \ref{fig:cartoons}(c).

The Griffiths phase further splits into two distinct regimes depending on whether $z<1$ or $z>1$. We denote the $\prare$ at which $z=1$ as $p_\text{D}$ (where $p_\text{D} < p_\text{SM}$ is a distinct error rate from the critical $p_\text{SM}$), at which point $\pfail$ is a constant greater than zero that is asymptotically independent of system size.\footnote{A constant $\pfail$ at threshold assumes that the power law does not have subleading multiplicative corrections. For example, $\sigma_\parallel \sim \log(L)/L^z$ would cause $\pfail$ at threshold to decay to $0$ with $L$. We do not expect any such multiplicative corrections.}
When $p_0^\text{2D} < p < p_\text{D}$, we have $0<z<1$, and the code is still decodable. 
Here, $\mathbb{P}_\text{fail}$ tends to zero as $L \to \infty$, since the cost $L^{1-z}$ diverges and $\pfail$ decays as a stretched exponential. 

On the other hand, $p > p_\text{D}$ means $z>1$, and the domain wall cost $L^{1-z}$ asymptotically vanishes, making $\pfail \to 1/2$ and the code nondecodable. Nevertheless, this regime is still in ordered phase of the stat mech model, due to the fact that \textit{transversal} domain walls (which are not related to $\mathbb{P}_\text{fail}$) continue to have a line tension $\sigma_\perp \sim \mathcal{O}(1)$, yielding a growing free energy cost; see Fig.~\ref{fig:columnar toric code} below.


When $p_{\text{rare}}$ is very large and above the critical disorder strength $p_{\rm SM}$ of the stat mech model, the system fails to order and the domain wall tension decays faster than any power law, giving a vanishing domain wall cost and $\pfail \approx 1/2$. This is captured by McCoy-Wu but is beyond the LRR approximation. We emphasize that the LRR nevertheless qualitatively captures the part of the phase diagram surrounding $p_D$, and only fails substantially above $p_D$. Note also that this discrepancy is small; the discrepancy is only in the qualitative description of how rapidly $\pfail$ approaches $1/2$ at sufficiently large $\prare$. 

\subsection{Numerical results}

\begin{figure}
    \centering
    \includegraphics[width = \linewidth]{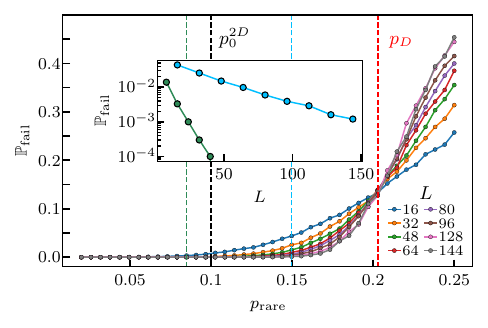}
    \caption{
    Threshold of the repetition code with time varying qubit error rates. For a length $L$ repetition code, we take $L$ rounds of measurements. 
    The threshold is at $\prare \approx 0.2$ and this is the point $p_{\rm D}$ in the phase diagram \ref{fig:cartoons}(c). In the inset, the median failure rate at $p_{\rm rare} = 0.08, 0.15$ (green, blue respectively). Though difficult to properly resolve the difference, when $p_{\rm rare} < p_0^{2D} \approx 0.1$, the failure rate is consistent with exponential decay and when $p_0^{2D} < p_{\rm rare} < p_{\rm D}$, the failure rate is consistent with stretched exponential decay in $L$. 
    }
    \label{fig:repcodenumerics}
\end{figure}

\begin{figure}
    \centering
    \includegraphics[width = \linewidth]{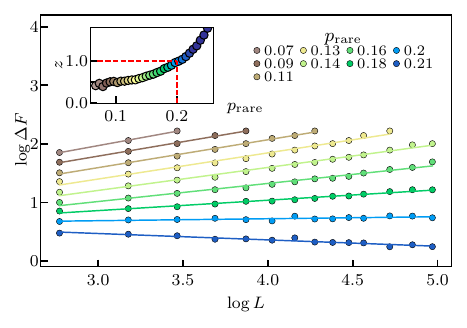}
    \caption{
    Stretched exponential power and defect cost scaling varies continuously throughout the phase. The inset shows the empirical value of $z$ obtained from the slopes of the lines in the panel as a function of $p_{\rm rare}$. Note that empirically, $z$ does not exactly equal $0$ for $p_{\rm rare} < p_0^{2D}$ (see Appendix \ref{sec:clean}) }
    \label{fig:varyingexponent}
\end{figure}

We now compare this predicted phase diagram to numerical simulations of the repetition code. For the numerics on the 1+1D repetition code, we take a length $L$ system with $L$ measurements.
We choose a uniform measurement error rate $p_{\rm meas} = 0.11$, which sets $p_0^{2D} = 0.10$.
We fix the bulk bitflip rate to be $p_{\rm bulk} = 0.02 < p_0^{2D}$, and vary the value of $p_{\rm rare}$.
We sample the rare regions according to the Bernoulli distribution Eq.~\eqref{eq:bernoulli}, where we take $\gamma = 1/3$ in both cases.

Our numerical results supporting the phase diagram are shown in Fig.~\ref{fig:repcodenumerics} and Fig. \ref{fig:varyingexponent}. In Fig.~\ref{fig:repcodenumerics}, we compute $\mathbb{P}_\text{fail}$ as a function of $p_\text{rare}$. We identify the decodable-to-nondecodable transition from the crossing of curves with different $L$ and find that it is clearly distinct from the threshold of the bulk system $p_0^{2D} \approx 0.1$\footnote{We also observe an empirical separation between the mean and typical (median) failure rate, not pictured in the main text, which we comment on in the Appendix \ref{sec:app_meanmedian}.}. In Fig.~\ref{fig:varyingexponent}, we depict the scaling of domain wall cost with system size. We see that there is an extended regime where the domain wall cost scales as a power-law with continuously varying exponent $1-z$, consistent with our prediction for $\prare$ within the Griffiths phase $(p_0^{2D}, p_{\rm SM})$. 

The inset of Fig.~\ref{fig:varyingexponent} shows the numerically extracted $z$ obtained from the linear fits in the main panel. We indeed find that it changes continuously with $p_\text{rare}$, dropping below 1 at $p_\text{D}$; however, we also observe that $z$ remains finite even when $p_\text{rare}$ is made smaller than the bulk threshold, contrary to our analytical argument. However, we note that our theoretical predictions neglected subleading contributions to the domain wall cost. In general, one might expect 
\begin{equation}
\Delta F \sim L^{1-z} + O(L^{\alpha})    
\end{equation}
where $L^{\alpha}$ with $\alpha < 1-z$ is a subleading term dependent on $\prare$. We do not attempt to compute such subleading corrections, but we note that these may cause significant finite-size effects, particularly in estimating $z$ numerically.
We view the persistence of $z>0$, even for $\prare < p_0^{2D}$ where we expect $z=0$, as an artifact of such subleading terms.\footnote{In Appendix~\ref{sec:clean}, we consider an $L \times L$ RBIM without correlated line disorder for comparison. There, we again fit $z$ in the same manner as in the inset of Fig.~\ref{fig:varyingexponent}, but we still find a $z$ that fails to reach $0$ even at small $p_{\rm bulk}$.}
To summarize, while our numerics on $z$ in Fig.~\ref{fig:varyingexponent} are suggestive, they are not yet conclusive.

\begin{figure}[h!]
    \centering
    \includegraphics[width=\linewidth]{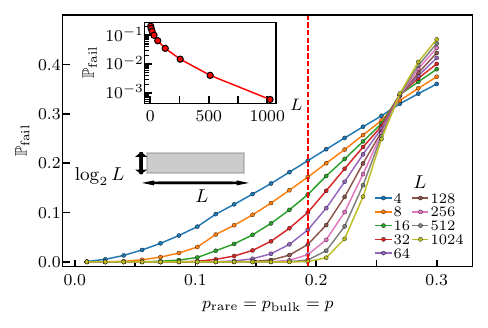}
    \caption{Threshold of a length $L$ repetition code with $\log_2 L$ measurements taken, with single error rate $p$. The defect cost only vanishes at a point $p_{\rm D}$ which is above $p_0^{2D}$. The inset shows a vertical cut through the plot at a point beneath the threshold where we observe the failure rate decaying as a stretched exponential in $L$.}
    \label{fig:logLrepcode}
\end{figure}

To further investigate our theoretical predictions, we directly investigate an $L \times \log_2 L$ system with uniform bitflip error rates $p_{\rm bulk} = \prare = p$. A uniform system of this size mimicks the largest rare region of an $L$ by $L$ heterogeneous system.
With numerical results shown in Fig.~\ref{fig:logLrepcode} for $\log_2 L \leq 10$, we confirm that a scale invariant point $p_{\rm D} > p_0^{2D}$ exists in the latter systems, and that failure rates below this point decay as stretched exponentials in $L$.
We note however that the numerical value of $p_{\rm D}$ is non-universal, so a direct comparison with the $L \times L$ heterogenous system cannot be made.

\begin{figure}[h!]
    \centering
    \includegraphics[width=\linewidth]{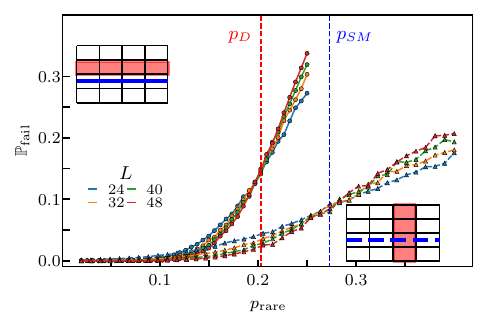}
    \caption{Different thresholds for horizontal (denoted by circles and solid lines) and vertical defects (denoted by triangles and dashed lines) in a repetition code. Insets show a diagram of the relevant defect and the rare regions. The vertical defect can studied via a $90^\circ$ rotation of the entire model, causing the correlation direction to now be in the temporal direction. The rare regions are shaded red and the defect is shown as a blue dashed line. The defect which is parallel to the correlation direction has a threshold at a $\prare \approx 0.2$. This is the point $p_{\rm D}$ in the phase diagram Fig \ref{fig:cartoons}(c). This data is a subset of \ref{fig:repcodenumerics}.
    The defect which is transverse to the correlation direction has a threshold at a $\prare \approx 0.27$. This is the point $p_{\rm SM}$ in the phase diagram. }
    \label{fig:columnar toric code}
\end{figure}

When $\prare$ is between $p_{\rm D}$ and $p_{\rm SM}$, the code is no longer decodable.
However, within this region, the overall phase of the RBIM is still ferromagnetic and the cost of the defect in the \textit{time direction} continues to grow.
This defect can be thought of as a ``temporal'' logical operator, relevant for fault-tolerant gates involving moving logical information in space~\cite{Gidney_2022}.
Our numerical results in Fig.~\ref{fig:columnar toric code} show that this logical 
has a higher threshold than when the logical was parallel to the correlation direction, confirming the phase diagram in Fig.~\ref{fig:cartoons}.
As the order-disorder transition occurs when defects in all directions cease to grow in free energy cost with system size, it is at $p_{\rm SM}$ where the system truly transitions to being paramagnetic.

Finally, we close this section by noting that $p_D$ will additionally depend upon the value of $\gamma$ used. Predictably, the code will be more tolerant to rare events if they are rarer and we show this explicitly in Appendix \ref{sec:app_distdep}.

\subsubsection{Additional numerical details for the repetition code}

Here we provide additional numerical details for Figs.~\ref{fig:repcodenumerics},~\ref{fig:varyingexponent}, \ref{fig:logLrepcode} and \ref{fig:columnar toric code}.
In Fig.~\ref{fig:repcodenumerics}, each point on each curve is determined from the median of $10^3$ error model realizations $\{ p_\alpha\}$. For each error model realization, $\pfail$ is determined from averaging over $10^4$ physical error configurations.
In Fig.~\ref{fig:varyingexponent}, $z$ is determined by a linear fit to $\log \Delta F$ (determined from Eq.~\eqref{eq:Delta_F_def}) vs $\log L$. The extracted value of $z$ should be viewed qualitatively. Due to potential subleading corrections which are present even in a system with uniform error rates (see Appendix ~\ref{sec:clean}), we do not see $z = 0$ for $\prare < p_0^{2D}$ as predicted. Furthermore, at low error rates and large system sizes, the median $\pfail$ is often 0; which is why the smallest values of $\prare$ are restricted to smaller system sizes. 
We note that techniques such as importance sampling (e.g. chapter 9 of Ref.~\cite{owen2023book}) may give greater access to this difficult regime of small $\prare$, and leave detailed exploration to future work.

In Fig.~\ref{fig:logLrepcode}, each point on each curve was determined from $10^6$ physical error configurations. 

In Fig.~\ref{fig:columnar toric code}, each point on each curve is determined from the median of $10^3$ error model realizations. $\pfail$ for each error model realization is determined from $10^4$ physical error configurations. To realize the transversal defect, we make the \textit{measurement error rate} a random variable in the spatial direction, and set the bit flip error rate to $0.11$.

\section{Planar Rare Regions for the 2+1D Toric Code \label{sec:planar}}




In this section we turn to time-dependent bitflip error rates in the 2D toric code.
They lead to rare regions in the random-plaquette gauge theory (RPGT) Eq.~\eqref{eq:rpgt} that are \textit{planar} and have infinite extent in the two spatial dimensions, see Fig.~\ref{fig:cartoons}(b).
Similarly to Sec.~\ref{sec:linear}, we draw bitflip error rates on different time steps from the Bernoulli distribution in Eq.~\eqref{eq:bernoulli}, and choose the time duration to be $T = L$.

We proceed as in Sec.~\ref{sec:linear}, first detailing our analytic predictions of the logical defect scaling for a quasi-2D system of dimensions $L \times L \times L_\perp$ with uniform error rates.
Treating this isolated system as a model of LRR, we then turn to the $L \times L \times L$ heterogeneous system with non-uniform error rates, where quantitative agreement between numerical results and analytic predictions are found. Our key finding is the absence of a decodable Griffiths phase: as soon as $p_\text{rare}$ exceeds the bulk threshold, and the rare regions are in the ``wrong'' phase, they make the decoder fail. As we will see, this is due to the 
2D nature
of the rare regions, which allows them to order by themselves.

\subsection{High temperature expansion within a clean planar LRR \label{sec:planar_HTE}}

As we have seen in Sec.~\ref{sec:linear}, \text{clean} models (where neither the random sign disorder or the spatial-temporal randomness in $\{K_\alpha\}$ is present) are useful for gaining intuition.

In parallel with Sec.~\ref{sec:assumption}, a high temperature expansion within a thin slab of dimensions $L \times L \times L_\perp$  yields
\begin{align} \label{eq:HTE_2+1D}
    \sigma_\parallel(L) \propto e^{-\alpha L_\perp \cdot L} \quad \Rightarrow \quad -\log \sigma_\parallel(L) \propto L \cdot L_\perp,
\end{align}
see details in Appendix \ref{sec:app_3dHTE}.
Once again, the LRR has width $L_\perp \propto \log L$, as in Eq.~\eqref{eq:LRR_height}.

As in Sec.~\ref{sec:assumption} and Eq.~\eqref{eq:sigma_exp_L_perp}, we perform this expansion for $K < K_c^{3D}$, where $K_c^{3D}$ is the critical coupling strength of the 3D uniform bulk system\footnote{Strictly speaking, the expansion is only valid for $K < K_c^{2D}(L_\perp)$, where $K_c^{2D}(L_\perp) < K_c^{3D}$ is the $L_\perp$-dependent critical coupling strength of the slab with height $L_\perp$.
Our statement here can be justified by noting that $K_c^{2D}(L_\perp)$ approaches $K_c^{3D}$ as $L_\perp$ increases, see Sec.~\ref{sec:planar_KW_dual},~\ref{sec:planar_crossover} for detailed discussions.}.
This result suggest that $\sigma_\parallel(L)$ vanishes exponentially with $L$ in the limit $L\to\infty$.
The defect cost is therefore $\Delta F \propto L^{1-z L}$ for some positive constant $z$, which vanishes in the thermodynamic limit when $K < K_c^{3D}$.
Translating this back to the decoding problem, an $L \times L$ toric code within time duration $T = L_\perp \propto \log L$ becomes immediately undecodable whenever $p > p_0^{3D}$, where $p_0^{3D}$ is the threshold of the 3D RPGT with uniform error rates.

Comparing Eq.~\eqref{eq:HTE_2+1D} with Eq.~\eqref{eq:sigma_exp_L_perp}, there is an extra multiplicative factor $L$ in $-\log \sigma_\parallel(L)$ due to the increased dimension of the rare region.
The strong vanishing of $\sigma_\parallel(L)$ in  Eq.~\eqref{eq:HTE_2+1D} suggests qualitative differences between linear and planar rare regions.




\subsection{Dual picture \label{sec:planar_KW_dual}}

Such qualitative differences can be more clearly appreciated in a dual picture. We summarize this here and explain this picture in depth in Appendix \ref{sec:app_kramerswannier}. Under a Kramers-Wannier duality, the linear and planar LRRs are described by quasi-1D and quasi-2D Ising models, respectively.
The defect line tension maps to the inverse dual correlation length;
the vanishing of the line tension (and therefore proliferation of line defects) corresponds to the ordering of the dual Ising models.
The difference in the scaling of $\sigma$ can be therefore attributed to whether the dimension of the rare region is below or above the lower critical dimension of the Ising model. 

In particular, for the planar rare regions, the critical \textit{dual} coupling strength $K_c^{2D, \ast}(L_\perp)$ of the $L \times L \times L_\perp$ dual Ising model is finite, and can be arbitrarily close to $K_c^{3D, \ast}$ of a 3D bulk Ising model, as $L_\perp \propto \log L$ can be arbitrarily large\footnote{In particular, under Kramers-Wannier duality, we have $K_c^{2D, \ast}(L_\perp) > K_c^{3D, \ast}$ for all $L_\perp < \infty$.
As $L_\perp \to \infty$, $K_c^{2D, \ast}(L_\perp)$ approaches $K_c^{3D, \ast}$ from above.
For any $K^\ast > K_c^{3D,\ast}$, we will also have $K^\ast > K_c^{2D, \ast}(L_\perp)$ for sufficiently large $L_\perp$.}.
Therefore, for any $K^\ast > K_c^{3D,\ast}$, the LRR will become ordered for sufficiently large $L_\perp \propto \log L$.

We can also understand these statements from the perspective of quantum models in one lower dimensions, related to the stat mech models via quantum-classical duality. We can relate the clean, $L \times L \times L_\perp$ classical problem to a quantum system on a $L \times L_\perp$ lattice\footnote{Note that in this case, the ``time'' direction corresponds to one of the spatial dimensions of the toric code, rather than to physical time.}. In this quantum model, the calculation of $\sigma(L)$ amounts to asking about the energy cost of inserting a point-like flux defect in the trivial paramagnetic phase (since we are dealing with rare regions). Using the quantum version of Kramers-Wannier duality, this is the same as the energy gap between the two symmetry broken ground states of a two-dimensional quantum Ising model in its \textit{ordered} phase, which is exponentially small in the \textit{volume} of the system, yielding the scaling in Eq.~\eqref{eq:HTE_2+1D}. The same argument could be used to deduce the scaling~\eqref{eq:sigma_exp_L_perp} in the case of 1D repetition code.

\subsection{Crossover scaling \label{sec:planar_crossover}}

The picture above offered by the Kramers-Wannier duality for the clean model suggest that when analyzing numerical data of $\pfail$ at $p_{\rm rare} \gtrsim p_0^{3D}$ of the toric code problem, we must take into account the proximity of $p_0^{3D}$ to the 2D critical point of the planar LRR.
We denote this $L_\perp$-dependent critical error rate as $p_0^{2D}(L_\perp)$, and we have $\lim_{L_\perp \to \infty} p_0^{2D}(L_\perp) = p_0^{3D}$.
As we increase $L_\perp$, we expect a crossover from the 2D RBIM transition to the 3D RPGT bulk transition.

To capture the finite $L_\perp$ crossover, we propose the following phenomenological two-parameter scaling function
\begin{align}\label{eq:scalingform1}
    \pfail(\prare, L, L_\perp) \approx \Phi\left[ (\prare-p_0^{3D}) \cdot L^{1/\nu_3}, L_\perp / L \right].
\end{align}
In a clean $L\times L \times L_\perp$ system without any disorder, such a crossover scaling form follows from Renormalization Group transformations.
We use the same form here despite the presence of disorder.
We expect the scaling function to be descriptive of $\pfail$ when both $L_\perp$ and $L$ are much larger than $1$, as usual in finite-size scaling.
Here, $\nu_3$ is the correlation length exponent for the 3D RPGT transition.
For any finite $L_\perp / L > 0$, this reduces to the conventional scaling function of the 3D RPGT with a nonzero aspect ratio.
The planar limit is $L_\perp / L \to 0$, where we expect to see critical scaling near a 2D critical point.
With these considerations, we find that 
\begin{align} \label{eq:p0_Lperp}
    p_0^{2D}(L_\perp) - p_0^{3D} = z_0 \cdot L_\perp^{-1/\nu_3},
\end{align}
where $z_0$ is a non-universal positive constant. See Appendix \ref{sec:app_crossover} for more detail.
We also find that at small $L_\perp / L$ the two-parameter scaling function $\Phi$ reduces to the following asymptotic form with a single parameter
\begin{align}\label{eq:scalingform2}
    \pfail &\stackrel{L_\perp / L \to 0}{=}
    \Phi^{2D} \left[ (\prare-p_0^{2D}(L_\perp)) \cdot L_\perp^{1/\nu_3} \cdot \left( L_\perp / L\right)^{-1/\nu_2} \right] \nonumber\\
    &=
    \Phi^{2D} \left[ \left( (\prare-p_0^{3D}) \cdot L_\perp^{1/\nu_3} - z_0 \right) \cdot \left( L_\perp / L\right)^{-1/\nu_2} \right].
\end{align}
Here, $\nu_2$ is the correlation length exponent for the 2D RBIM transition, and $\Phi^{2D}$ is the corresponding \textit{universal} scaling function.
Note that for any finite $L_\perp$, this function reduces to the standard 2D scaling form with the argument $\propto (\prare-p_0^{2D}(L_\perp)) \cdot L^{1/\nu_2}$.

\subsection{Predicted phase diagram}

With these, and the assumptions that the logical failure of the code is dominated by the LRR (see Sec.~\ref{sec:assumption}),
we now discuss the predicted phase digram in Fig.~\ref{fig:cartoons}(d).
Recall that we choose $p_{\rm bulk} < p_0^{3D}$, and vary $\prare$ across $p_0^{3D}$.

When $\prare < p_0^{3D}$, the entire system has error rate below $p_0^{3D}$, and the code is in its decodable phase.
Correspondingly, the RPGT is in its deconfined phase, where the conventional scaling $- \log \pfail \propto L$ holds.

When $\prare > p_0^{3D}$, we predicted in Sec.~\ref{sec:planar_HTE},~\ref{sec:planar_KW_dual} that the code becomes undecodable in the thermodynamic limit, i.e. $\lim_{L\to\infty} \pfail \to 3/4$ for any $p_{\rm bulk}$.
In Fig.~\ref{fig:cartoons}(d), we highlight the critical point $p_{\rm D}(L) \equiv p_0^{2D}(L_\perp \propto \log L)$ of the planar LRR, which approaches $p_0^{3D}$ from above as $L \to \infty$.
For $\prare$ between $p_0^{3D}$ and $p_{\rm D}(L)$, the system is in the ``mesoscopic crossover regime'' (dashed green region), whose extent shrinks with increasing $L$.
Within this regime, we expect $\pfail$ to be suppressed with increasing $L$ for $L$ below a crossover length scale $\xi_{\rm crossover} \propto (\prare - p_0^{3D})^{-\nu_3}$.
Above the crossover length scale, $\pfail$ grows with $L$, and eventually saturates to $3/4$ in the infinite $L$ limit.\footnote{{ When increasing $L$, the logical failure rate $\pfail$ is minimized at $L \sim \xi_{\rm crossover}$ before increasing again, see Fig.~\ref{fig:tcdisordered} and~\ref{fig:logTC}. This trend is similar to an ``error floor'' observed in recent QEC experiments~\cite{google2023suppressing, acharya2024quantum}, reached when the code patch is comparable to a finite crossover length. Our results can potentially provide a heuristic explanation of this experimental observation, which is an interesting direction left for furture work.} }
Therefore, a non-monotonic dependence of $\pfail$ on $L$ is expected.
All these phenomenology should be captured by the scaling function Eq.~\eqref{eq:scalingform1} for sufficiently large $L$, $L_\perp$, and for $\prare$ sufficiently close to $p_0^{3D}$.

Similar to the 1+1D repetition code, there is again a range of values of $\prare$ between $p_D(L)$ and $p_{\rm SM}$, where the rare regions are disordered but the line tension $\sigma_\perp$ of the transveral defect (perpendicular to the planar rare regions) is still finite. Due to the presence of the disordered rare regions, we call this regime a Griffiths phase but unlike the 1+1D repetition code, it is non-decodable.

\subsection{Numerical results}

\begin{figure}[t!] 
    \centering
    \includegraphics[width=\linewidth]{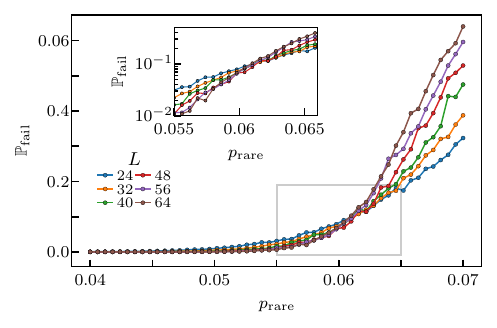}
    
        \caption{$\pfail$ vs $p_{\rm rare}$ for an $L \times L$ toric code with time varying qubit error rates and $L$ rounds of measurements.
        A clear crossing is not observed in this data. In the inset, we zoom in on the ``crossing" region of the plot (marked by the gray box) and see that on a log scale, a single crossing point cannot be readily distinguished and the data is consistent with our theoretical prediction of a backwards drifting crossing. This is the backwards drifting $p_D(L)$ in Fig. \ref{fig:cartoons}(d).}
    \label{fig:tcdisordered}
\end{figure}

\begin{figure}[t!] 
    \centering

    \includegraphics[width=\linewidth]{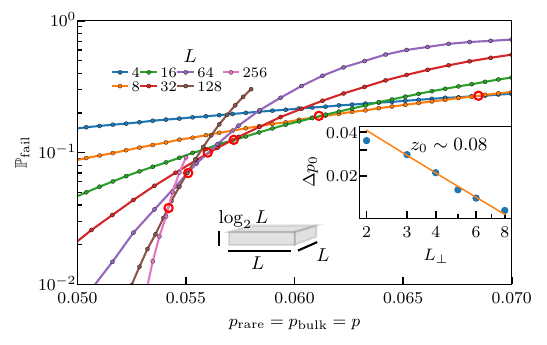}
    
        \caption{The logical failure rate of a $L \times L$ toric code under $\log_2 L$ rounds of measurements, with uniform bit flip error rate and measurement error rate ($0.01$).
        There is no finite size crossing, unlike with the repetition code. Thus, there is no thermodynamic phase transition at $p_{\rm D} > p_0^{3D}$. Instead the crossing point drifts backwards with increasing $L$.
        The critical error rate of each system depends on the short dimension $L_\perp$.
        In the inset, we plot $\Delta p_0 = p_0^{2D}(L_\perp) - p_0^{3D}$ as a function of $L_\perp$. For the values of $L$ shown in the main panel, we set $L_\perp = \log_2 L$ and find a critical error rate for each $L_\perp$. We find  the expected power law dependence Eq.~\eqref{eq:p0_Lperp} with fitting parameter $z_0 = 0.08$ and $\nu_3 = 1$.}
    \label{fig:logTC}
\end{figure}



We calculate $\pfail$ for both an $L\times L \times L$ heterogeneous system with rare regions, and an $L\times L \times \log_2 L$ isolated system with uniform error rates.
In both cases, we choose a uniform measurement error rate $p_{\rm meas} = 0.01$, which sets $p_0^{3D} = 0.045$.
For the $L \times L \times L$ system, we choose $p_{\rm bulk} = 0.01 < p_0^{3D}$, and generate the rare regions according to Eq.~\eqref{eq:bernoulli} with $\gamma = 1/3$.

Our numerical results are shown in Fig.~\ref{fig:tcdisordered} and~\ref{fig:logTC}.
In Fig.~\ref{fig:tcdisordered} for the $L\times L \times L$ system, we find no scale-invariant point $p_{\rm D}$ for the median $\pfail$.
Instead, we find a crossover regime with a non-monotonic dependence of $\pfail$ on $L$.
This is consistent with our expectations from the discussions above.
In Fig.~\ref{fig:logTC} for the isolated $L\times L \times L_\perp$ system, these predictions are more clearly borne out.
When setting $L_\perp = \log_2 L$, we observe a clear downward drift of $p_D(L)$ towards $p_0^{3D}$ with increasing $L$, as well as the non-monotonic dependence of $\pfail$ on $L$.
We also extract $p_0^{2D}(L_\perp)$ at fixed finite values of $L_\perp$.
The results can be fitted well to Eq.~\eqref{eq:p0_Lperp}, see inset of Fig.~\ref{fig:logTC}.

\begin{figure}
    \centering
    \includegraphics[width=\linewidth]{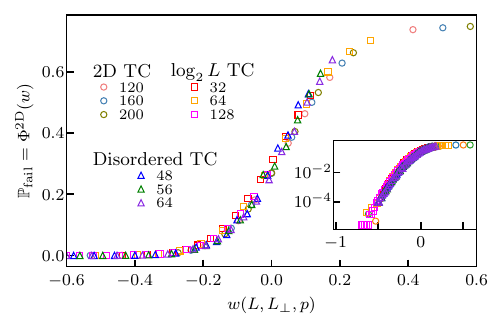}
    \caption{Scaling collapse for $\pfail$ in the 2D toric code with $L$ rounds of measurements and temporally-varying error rates (triangles), the 2D toric code with $\log_2 L$ rounds of measurements and uniform error rates (squares), and the 2D toric code with a single round of perfect measurement (circles).
    By $w(L, L_\perp, p)$ we denote the argument of the single-parameter scaling function Eq.~\eqref{eq:scalingform2}.
    With appropriate rescaling, these collapses lie on top of each other, as their functional forms are all given by $\Phi^{2D}$, see Eq.~\eqref{eq:scalingform2}.
    In the inset, we plot the same data on a log scale for a better resolution of points at small $\pfail$.}
    \label{fig:collapse}
\end{figure}

Furthermore, the single-parameter scaling function Eq.~\eqref{eq:scalingform2} is relevant as $L_\perp \ll L$, and can be directly tested.
In Fig.~\ref{fig:collapse}(triangles) we show the data collapse of the heterogeneous $L\times L\times L$ system, and find good agreement with Eq.~\eqref{eq:scalingform2} when setting $L_\perp \propto \log L$ in its parameter.
In Fig.~\ref{fig:collapse}(squares), collapse of data from the $L\times L \times L_\perp$ system is shown, where agreements are again found.

Finally, the functional form of $\Phi^{2D}$ in Eq.~\eqref{eq:scalingform2} is universal and thus can be extracted independently by decoding the 2D toric code with perfect measurements.
These results (Fig.~\ref{fig:collapse}(circles)), when plotted together with the numerical scaling collapses from the previous two numerical experiments, lie on top of each other (up to a constant rescaling of their parameters), see Fig.~\ref{fig:collapse}.
This strongly supports the crossover scaling picture provided by Eqs.~(\ref{eq:scalingform1},\ref{eq:scalingform2}).
In particular, the data collapse is strong evidence for our theoretical prediction that the 2D scaling function, modulo a rescaling of its argument, describes the failure probability close to threshold even in the (2+1)D planar-disordered toric code described by a 3D stat mech model.

To summarize, in the toric code, whenever $p_{\rm rare} > p_0^{3D}$, planar rare regions render decoding entirely impossible in the thermodynamic limit\footnote{As a consistency check, this should follow directly from Eq.~\eqref{eq:scalingform2}. Taking $L_\perp \propto \log L$, we indeed find $\lim_{L\to\infty} \pfail = 3/4$ for any $p > p_0^{3D}$.}. However, a transverse defect perpendicular to the planar rare regions may still have a diverging $\Delta F$. Indeed, the defect cost will diverge whenever the transvserse correlation length in the dual Ising model is finite. It is known~\cite{Vojta_2006} that one can have a spontaneous magnetization without a diverging transverse correlation length in 3d Ising models with planar disorder, corresponding in the primal gauge theory to a non-decodable region that has a diverging transvserse defect cost. These dualities are complicated by the presence of quenched random sign disorder, which we have largely neglected in our discussion of dualities. We do find such a region via numerical results on small systems $L \leq 18$ in Appendix \ref{sec:app_transversal}, and we find a transition in the transverse defect cost at a $p_{\rm SM} > p_0^{3D}$. We leave a more thorough investigation of this transition to future work.


\subsubsection{Additional Numerical Details}

Here we provide additional numerical details for Fig. \ref{fig:tcdisordered} and Fig. \ref{fig:collapse}. In Fig \ref{fig:tcdisordered}, each point on each curve is determined from the median of $10^3$ error model realizations and $\pfail$ for each error model realization is determined from $10^4$ physical error configurations. For Fig. \ref{fig:logTC}, for $L = 4,8,16,32$, each point is determined from $10^5$ physical error configurations. For $L = 64,128,256$ we use $10^4$ physical error configurations. In the inset, for each value of $L_\perp$, we obtain the critical error rate from system sizes of $L = 24$ to $L = 64$. For each system size, we obtain $\pfail$ from $10^5$ physical error configurations. 

In Fig. \ref{fig:collapse}, we plot the same data from Fig. \ref{fig:tcdisordered} along with additional data collected for a toric code with no measurement errors.  In all data collapses, we take $\nu_3 = 1$ and $\nu_2 = 1.5$.
We use $z_0 = 0.08$ for the planar-disordered 2+1D toric code and $z_0 = 0.09$ for the $\log_2 L$ 2+1D toric code. 
$z_0$ is not universal, and it is reasonable to expect different $z_0$ values for these two systems.
For the 2D toric code with no meausrement errors, we use $w = (p - p_0^{2D})\cdot L^{1/\nu_2}$ for the argument of the scaling form, where $p_0^{2D} \approx 0.103$~\cite{Wang_2003}.
A slight $x$ axis rescaling was also necessary for good collapse. For the $\log_2 L$ toric code, we rescale the $x$ axis by 1.1 and for the disordered toric code, we rescale by 1.2.

\section{Conclusions}

In this work we analyzed the performance of topological quantum codes in the presence of non-uniform error rates that are long range correlated.
We find that rare events with increased error rates (above the bulk threshold) can have dramatic effects on the code performance.
We point out crucial differences between linear and planar rare regions: the former lead to a new decodable phase for the 1D repetition code where the logical failure rate scales as a stretched exponential in the code distance, while the latter make decoding immediately impossible for the 2D toric code, as soon as the error rate in the rare region exceeds the global decoding threshold
We expect these analyses to be more broadly applicable to other topological codes with point-like excitations.

{Our analysis can be immediately extended to the case with fabrication errors, leading to parts of the code patch which have a larger than bulk error rate.
They will result in linear rare regions parallel to the temporal direction, for both the 1D repetition code and the 2D toric code.
As such, the dominant errors in this setup are time-like errors which can lead to logical failure in stability experiments and lattice surgery~\cite{Gidney_2022, bjbrown2024anyoncondensation}.
Our analysis based upon high-temperature expansion (Appendix~\ref{sec:linetensions}) and Kramers-Wannier duality (Appendix~\ref{sec:app_kramerswannier}) for linear rare regions implies that that a decodable stretched exponential phase is present for both codes, see the relevant Appendices for details.}

While we focused on toy error models for simplicity, the asymptotic scaling thus predicted are expected to be universal, and can hold for more realistic error models. 
For instance, fabrication errors may lead to qubits with different fixed error rates and this may be better modeled by a noise distribution other than Bernoulli~\cite{Strikis_2023, aasen2023faulttolerant}.
For cosmic ray events, an error model incorporating the relaxation time of the qubits can be considered.

While current quantum error correction (QEC) experiments are restricted to small code distances, we expect our results to be descriptive of future experiments when they approach the scaling limit.
More broadly, QEC experiments provide new platforms and new motivations for exploring disorder physics.
This work provides first explorations in this direction, and in future work it would be interesting to consider other models, including those in (generalized) gauge theories which are themselves under-motivated in solid state systems.

~

\section*{Acknowledgements}

We acknowledge helpful discussions with Aditya Mahadevan, Arpit Dua, David Huse, Chaitanya Murthy and especially Akshat Pandey.
We thank Hengyun Zhou for helpful discussions and for bringing Ref.~\cite{tan2024resilience} to our attention.

A.S. acknowledges support from the National Science Foundation Graduate Research Fellowship. N.O.D. acknowledges support
from the ARCS Foundation for ARCS Scholar funding.
Y.L. was supported in part by the Gordon and Betty Moore Foundation’s EPiQS Initiative through Grant GBMF8686, and in part by the US Department of Energy, Office of Science under Award No DE-SC0019380.
Y.L. and T.R. were supported in part by the Stanford Q-FARM Bloch Postdoctoral Fellowship in Quantum Science and Engineering. 
V.K. acknowledges support from the Office of Naval Research Young Investigator Program (ONR YIP) under Award Number N00014-24-1-2098, the Alfred P. Sloan Foundation through a Sloan Research Fellowship and the Packard Foundation through a Packard Fellowship in Science and Engineering.
Numerical simulations were performed on Stanford Research Computing Center's Sherlock cluster.

\appendix

\section{{Rare Regions and Griffiths Effects}}\label{sec:mccoywu}
{
In this section we aim to provide a brief overview of rare effects that can occur due to extended regions of disorder in statistical mechanics models. For a more thorough introduction we refer the reader to the review article~\cite{Vojta_2006}.}

{
Quenched disorder is used to model systems in which impurities and inhomegenity in the system properties fluctuates on a time scale which is much slower than the fluctuation of the degrees of freedom. This type of disorder leads one to consider systems in which there is a finite density of spatial inhomogenity making straightforward analysis difficulty. For example, disorder may lead to a finite density of antiferromagnetic interactions in a system which otherwise has completely ferromagnetic interactions. It is natural to ask whether this truly matters. For example, if the disorder is weak and uncorrelated and if the physics is dominated by long wavelength properties then indeed one might observe a strong agreement between the disordered system and its clean analogue. On the other hand, if the disorder is strong, it can lead to an outsized affect in the thermodynamics.}

{
A basic question is then, can the disorder affect the critical temperature? If it does, then it is possible for different parts of the system to order/disorder independently, which may destroy a sharp phase transition in the system. A heuristic criterion provided by Harris informs in which systems this occurs. Assuming the existence of a true critical temperature, if one considers a bulk system at some temperature away from this temperature, then the system may be subdivided into "boxes" of size $\sim\mathcal{O}(\xi)$ where $\xi$ is the correlation length. Each of these boxes can be thought of as a finite size disorder realization. If each of these boxes is treated as an individual system, then there is a temperature at which the correlation length for a specific box is of order the size of the box. This can be thought of as a critical temperature. If one considers the distribution of these critical temperatures, as the global temperature is tuned towards the true critical temperature, if the distribution of these critical temperatures is smaller than the distance to the true critical temperature, then there exists a sharp phase transition. Said another way, as the correlation length increases, the distribution of critical temperatures sharpens around the true critical temperature. This phenomen occurs as long as $d\nu > 2$, where $d$ is the dimensionality of the system and $\nu$ is a critical exponent. }

{
It is when the Harris criterion fails to be satisfied that rare region and Griffiths effects may occur. Precisely due to the fact that individual rare regions may undergo phase transitions independently, atypical configurations play an outsized role in the thermodynamics. Indeed this is what occurs in the McCoy-Wu model, referenced in the main text \cite{PhysRev.176.631}. In the McCoy-Wu model, the rare regions are extended in one-dimension and so can be thought of as disordered while the remainder of the system is ordered. Another example of where this occurs is in a 3D Ising model with planar disorder \cite{vojta2004planar}. In this model, individual planes of the 3D Ising model can order and disorder on their, destroying the sharp phase transition. }

{
For readers familiar with quantum phase transitions, much of this discussion applies there as well. For instance, the random transverse field Ising model (rTFIM) (which is the quantum analogue of the McCoy-Wu model) contains Griffiths effects due to the fact that the disorder of that model has an extended dimension in \textit{imaginary time} \cite{PhysRev.176.631, PhysRev.188.982}. As a result, individual sections of the Ising chain order and disorder independently. In rTFIM  In this Griffiths phase, the energy of a domain wall scales as $S \sim \frac{1}{L^z}$ where $L$ is the length of the chain and $z$ is a dynamical exponent which depends on the difference between the average of the logarithms of the transverse field distribution and the coupling distribution. The domain wall energy can be identically defined as the energy difference between placing the chain on periodic and anti-periodic boundary conditions. In contrast, in a clean TFIM, this value will be a constant. The energy of this domain wall in the 1D model can be mapped to the vertical domain wall free energy in the classical McCoy-Wu model, yielding a sublinear scaling of the vertical domain wall free energy \cite{PhysRevB.51.6411}. }

{
Similarly, the quantum analogue of the planar disordered gauge theory we discuss in the main text is a 2+1D quantum gauge theory with strip disorder. Each of the strips in such a model are slab-like due to the imaginary time direction. That these strips can disorder individually means that the gap to the relevant defect (one flipped plaquette) decays exponentially in the area of the strip, which is the same as the defect tension scaling in the 3D gauge theory with slab defects presented in the main text.}

\section{Minimum Weight Perfect Matching}\label{sec:matching}

Throughout this work, we use minimum weight perfect matching (MWPM) for decoding.
MWPM finds the minimum weight error consistent with the observed syndrome.
As the weight of an error is associated with an energy, MWPM neglects entropic contributions and can be thought of as a zero temperature decoder. 

MWPM works as follows.
First, a \textit{decoding graph} is defined such that every node of the graph corresponds to a check of the code, and every edge in the graph represents a local error that connects all checks triggered by this error.
We assume that each edge connects at most two checks (i.e. no hyperedges), which is the case for all the model considered in this paper.
Second, a syndrome measurement marks which of the checks were triggered and those nodes are highlighted on the decoding graph.
Next, the decoder works to pair up each of the highlighted nodes on the check graph with the minimum number of edges.
From such a minimum weight matching, a proposed correction operation can be obtained.
The pairing can be carried out efficiently in $O(n^3)$ time, where $n$ is the number of nodes on the decoding graph~\cite{PhysRevX.9.021041, edmonds}.

To account for non-uniform error rates, weights can be assigned to edges based upon the rate of the corresponding error.
If error $a$ has error rate $p_a$, then its respective edge on the decoding graph has weight $\ln\frac{1-p_a}{p_a}$. 
Throughout this work, we supply the decoder with the information about the non-uniform error rates.

\subsection{Repetition Code and Toric Code with Uniform Error Rates}\label{sec:clean}

\begin{figure}[h!]
    \centering
    \includegraphics[width = \linewidth]{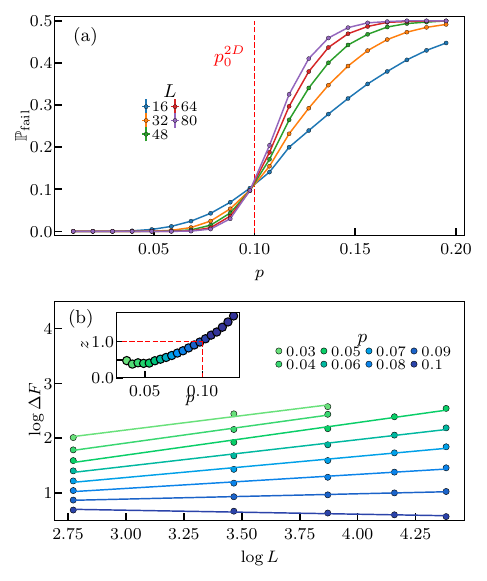}
    \caption{(a) Threshold of repetition code with uniform bit flip error rates. The bit flip error rate at the finite size crossing is $p_0^{2D}$. Each point was calculated from $10^6$ physical error configurations. The measurement error rate was set to 0.11. (b): Defect cost scaling extracted from (a), where the defect cost form is assumed to be $L^{1-z}$. One expects to see $z = 0$ throughout the entire decodable phase but due to subleading corrections, this is not empirically observed.  
    }
    \label{fig:cleanrepcode}
\end{figure}

\begin{figure}[h!]
    \centering
    \includegraphics[width = \linewidth]{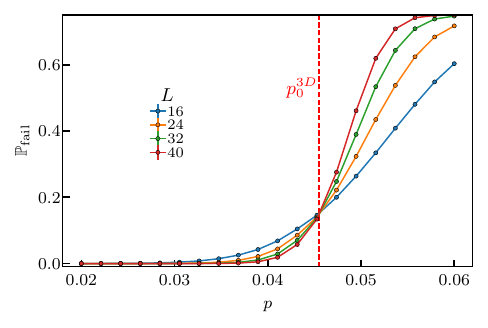}
    \caption{Threshold of toric code with uniform bit flip error rates. The bit flip error rate at the finite size crossing is $p_0^{3D}$. This point was found to be $p \sim 0.045$. Each point was calculated from $10^5$ physical error configurations. The measurement error rate was set to 0.01. 
    }
    \label{fig:cleantoriccode}
\end{figure}

Here we provide baseline numerics using MWPM for the repetition code and the toric code. We utilize a uniform error rate in both systems in order to obtain $p_0^{2D}$ and $p_0^{3D}$. This data is shown in Fig. \ref{fig:cleanrepcode}(a) and Fig. \ref{fig:cleantoriccode} respectively. The values we obtain are consistent with those found in the literature \cite{Dennis_2002}.

For the repetition code, we also calculate the defect cost similar to the procedure used in Fig. \ref{fig:varyingexponent}. This is shown in Fig. \ref{fig:cleanrepcode}(b). The defect cost is expected to scale purely exponentially (i.e. $z = 0$). However, due to subleading corrections in the defect cost scaling, the empirical value of $z$ is slightly greater than 0. Furthermore, the error rates shown here are close enough to the threshold that we cannot rule out contributions to the scaling form from critical scaling. This is to be contrasted with the case of the rare-event repetition code in the main text where the nonzero empirical value of $z$ for $\prare > p_0^{2D}$ is theoretically understood to be coming from a sublinear leading term, and we see its effects reasonably far from the threshold.

\subsection{Griffiths physics for MWPM}\label{sec:app_MWPM_griffiths}

We used free energies of defects to inform our theoretical treatment of success probabilities. This picture holds for the maximum likelihood (ML) decoder, where success probabilities can be reframed in terms of free energies of defects in statistical mechanical models with couplings $K=\beta J$ set in terms of the error rates by ``Nishimori conditions."
For MWPM, $\pfail$ is actually set by the probability that introducing a defect lowers the energy of the system. Quantities that depend on details of energies probe zero temperature physics, which can be in a different phase from the finite-temperature phase probed by the ML decoder.
However, we argue that the relevant Griffiths physics at play does not change much from the picture for the ML decoder.

For specificity, consider the case of the RBIM. Here, the defect is a domain wall formed by flipping the sign of a column of couplings. For sufficiently large disorder strengths at zero temperature, there is a phase transition from the ferromagnet to a spin glass phase. This spin glass phase will control the physics of the ``wrong phase" spatial regions in the Griffiths phase.
Assume that $\pfail$ is largely set by the largest rare region living in the wrong phase. In an $L$ by $L$ system, this largest rare region has size $L_\perp$ by $L$ with $L_\perp$ scaling as $\ln(L)$. The corresponding $\pfail$ for MWPM on a model with dimensions of the rare region will be bounded below by that of the maximal likelihood decoder on the same region. 

However, the maximal likelihood decoder has its failure probability controlled by free energy differences, and the free energy difference in the paramagnetic phase is exponentially suppressed in system size as $a L e^{-b L_\perp}$ for some $a$ and $b$ that are functions of error rates but are independent of system size. This gives $\pfail \sim 1/(1+e^{a L e^{-b L_\perp}})$ for the ML decoder whenever the error rate is sufficient to enter the paramagnetic phase. Note again that $\pfail$ for the MWPM decoder is also bounded below by this quantity; though this is not a proof, we believe it plausible that $\pfail$ for the MWPM decoder in the spin glass behaves in a similar manner to that of the ML decoder in the paramagnet. That is, we believe $\pfail \sim 1/(1+e^{a' L e^{-b' L_\perp}})$ for some new functions $a',b'$ of the error rates. Furthermore, we expect this functional form to hold through the whole spin glass phase, not only where the error rate is sufficient for the better ML decoder to leave the ferromagnetic phase. Under this assumption, the phases of $\pfail$ in Fig.~\ref{fig:cartoons}c) remain unchanged for the MWPM decoder, even if the phase boundaries shift relative to the ML decoder. 

An analogous argument holds for MWPM in the 3D gauge theory.

\subsection{Mean-Median Separation}\label{sec:app_meanmedian}

\begin{figure}
    \centering
    \includegraphics[width=\linewidth]{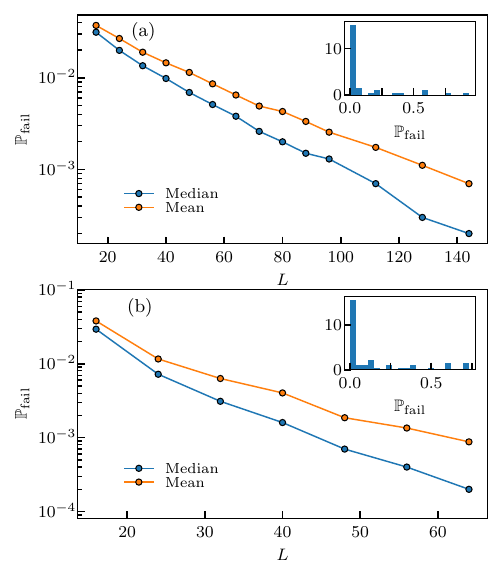}
    \caption{Mean-median separation of $\pfail$ for repetition code (a) and toric code (b). For the repetition code, $p_{\rm bulk} = 0.02, \prare = 0.14$ and the measurement error rate is $0.11$. Data was calculated from $10^3$ error model realizations and $10^4$ physical error configurations per disorder realization. For the toric code, $p_{\rm bulk} = 0.01, \prare = 0.052$ and the measurement error rate is $0.01$. Data was calculated from $10^3$ error model realizations and $10^4$ physical error configurations per error model realization. In the insets, the distributions of $\pfail$ across error model realizations for the largest system size are shown.}
    \label{fig:meanmedian}
\end{figure}

In the main text, we noted that our numerical results yielded a separation between the mean and median logical failure rates as calculated from the MWPM decoder. This phenomenon is shown in Fig. \ref{fig:meanmedian}. When there is uncorrelated or short range correlated disorder, this $\pfail$ distribution is expected to be approximately Gaussian. We find that this is emphatically not the case for long range correlations. There are several sources of randomness that could lead to asymptotically different scaling forms for the two quantities. For instance, the largest rare region in a given realization could be larger than its typical size by a big $\mathcal{O}(1)$ constant times $\log(L)$, which could enhance the failure rate to be $\mathcal{O}(1)$ instead of decaying. Those rare disorder realizations, which occur with probability $\sim 1/{\text{poly}(L)}$, would skew the mean failure probability to be at least $\sim 1/{\text{poly}(L)}$ rather than stretched exponential. Such uncommon realizations will cause the mean to scale differently than the median. Though we have theoretical reason to suspect that the mean and median should scale differently (power-law versus stretched exponential), we do not resolve this asymptotically different scaling in the numerics.

\subsection{Distribution Dependence} \label{sec:app_distdep}

\begin{figure}
    \centering
    \includegraphics[width=\linewidth]{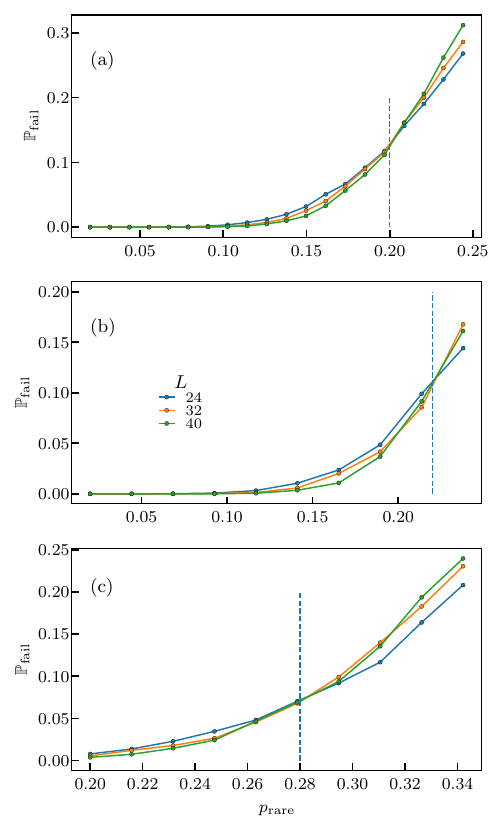}
    \caption{Distribution dependence of threshold of repetition code with Bernoulli distributed error rates. Here, $\gamma = 1/3,1/4,$ and $1/9$ in the (a), (b), and (c) panels respectively. We observe that the finite size crossing in the failure rate drifts rightward as the probability of a rare event decreases and note that this is consistent with the fact that the size of the typical largest rare region should have a weak power law dependence on the probability of the rare event. Curves are the median from 500 disorder realizations each and $10^4$ physical error configurations per disorder realization.  The measurement error rate was $0.11$ and $p_{\rm bulk} = 0.02$.}
    \label{fig:distdependence}
\end{figure}

In Fig. \ref{fig:distdependence},  we present data on how $p_D$ for the repetition code changes with $\gamma$. We predictably find that $p_D$ increases with decreasing $\gamma$.

\section{Derivation of Defect Tensions}\label{sec:linetensions}

Here, we will derive the defect tension for both the Ising model and the plaquette gauge theory using low and high temperature expansions. These derivations will be asymptotic in nature. Though the models discussed in the main text are disordered and contain couplings with both sign and magnitude disorder, we expect that the bulk phase properties to have the same asymptotic scaling behavior as the clean models.  

\subsection{2D Ising Model}\label{sec:app_2dHTE}

In a ferromagnet, the defect cost scales as $L$ with a $\mathcal{O}(1)$ line tension. We can directly calculate this from the low temperature expansion. At lowest order in $e^{-K}$, the partition function of an $L$ by $L_\perp$ clean Ising system is $Z \sim 2 e^{2K L L_\perp}$. On introducing a domain wall via antiperiodic boundary conditions along the length $L$ direction, the lowest order contribution is $Z_{DW} \sim 2L_\perp e^{2 K L L_\perp -2KL}$. This contribution comes from the $L_\perp$ locations that the shortest domain wall of length $L$ can be placed. The definition of the domain wall free energy (up to a multiplicative factor of $\beta$) is $\Delta F = \log(Z) - \log(Z_{DW})$, giving $\Delta F \sim 2 K L - \log(L_\perp) \sim 2KL$ and a domain wall tension of $2K$. Higher-order contributions renormalize the domain wall tension of $2K$ via a series in $e^{-K}$.

In a paramagnet, the defect cost should scale as $L e^{-zL_\perp}$. We may calculate this from the high temperature expansion, which sums over weighted loops on the real lattice. Recall the high temperature expansion of the Ising model:
\begin{align}
    Z = (\cosh K)^{2N} \sum_{ \{ \sigma_i \} } \prod_{i,\nu} \left( 1 + (\tanh K) \sigma_i \sigma_{i+\nu} \right)
\end{align}
where $\nu$ is a lattice translation. Since $\sigma_i^2 = 1$ and $\sum_{\sigma_i = \pm 1} \sigma_i = 0$, only terms which contain no factors of $\sigma_i$ can contribute to the partition function. These terms correspond to closed loops built out of bonds. The first few terms of the partition function series are
 \begin{align}
     Z = (\cosh K)^{2N} \left(1 + L L_\perp (\tanh K)^4 + 2 LL_\perp (\tanh K)^6 + \ldots \right)
 \end{align}
The effect of anti-periodic boundary conditions (i.e. flipping the sign of the couplings on some column) is to put a $-$ sign on some of the loops. Note that any contractible loop must contain an even number of links on the antiferromagnetic column.
Thus, the first nontrivial contribution to
\begin{equation}
    \Delta F = \log(Z) - \log(Z_{DW})
\end{equation}
will be from the shortest-length non-contractible loops that run perpendicular to the antiferromagnetic column. These loops have length $L_\perp$, and there are $L$ of them. 
Since such loops carry opposite signs between $Z$ and $Z_{DW}$, we have
\begin{equation}
\label{eq:domain_wall_tension_expansion_appendixB}
    \Delta F \sim 2 L (\tanh K)^{L_\perp}
\end{equation}
with manifest exponential decay in $L_\perp$.

\subsection{3D Gauge Theory}\label{sec:app_3dHTE}

Magnetic flux tubes are the defects of the $\mathbb{Z}_2$ lattice gauge theory stat mech model corresponding to the 2+1D toric code. In the low temperature phase of the clean gauge theory, we may obtain the cost of this defect through a low temperature expansion. This cost comes from the free energy difference of flipping the sign of the coupling of a column of plaquettes. The defect cost is
\begin{align}
    \Delta F &= F_{\rm flipped\ column} - F_{\rm no\ flipped\ column} 
    \\&= -\log \frac{{Z}_{\rm flipped\ column}}{{Z}_{\rm no\ flipped\ column}}.
\end{align}

Similar to the 2D Ising model, at lowest order in $e^{-K}$, the partition function of a clean $L_x \times L_y \times L_z$ system is $Z_{\rm no\ flipped\ column} \sim 4e^{3K L_x L_y L_z}$. Upon fixing a flux tube by flipping a the sign of row of plaquettes transverse to the $xy$ plane, the lowest order contribution to $Z_{\rm flipped column}$ is $\sim 4L_x L_y e^{K(3L_x L_y L_z - 4L_z}$. This form is the result of the column of antiferromagnetic bonds lowering the energy of the ground state configuration. There are $L_x L_y$ places to put this column. The flux tube cost becomes $\Delta F = \log (Z_{\rm no flipped column}) - \log (Z_{\rm flipped column})$ again up to a multiplicative constant of $\beta$. This reduces to $\Delta F \sim 4KL_z - \log L_x L_y \sim 4KL_z$. The flux tube tension is an $\mathcal{O}(1)$ number, $4K$ and again, higher order contributions renormalize the tension in powers of $e^{-K}$.

Next, we calculate how the defect cost should scale in the high temperature phase. The first non-unity term in the high temperature expansion are closed surfaces of area 6. However, any closed surface which is contractible will always contain even numbers of plaquettes from the tube, similar to the 2D case. Therefore, the first term which is different between the $Z_{\rm flipped\ column}$ and $Z_{\rm no\ flipped\ column}$ will be the product of all the plaquettes within a plane pierced by the tube. This occurs at order $L_xL_y$. We obtain the defect cost in the high temperature phase to be
\begin{equation}
    \Delta F \sim 2 L_z (\tanh {K})^{L_xL_y}
\end{equation}
with manifest exponential decay.
We see that the defect cost decreases exponentially in the area of the plane within the high temperature phase.

\section{Discussion of the phase diagrams with Kramers-Wannier duality}\label{sec:app_kramerswannier}

In this Appendix, we provide additional discussion of the phase diagrams in Fig.~\ref{fig:cartoons}. We emphasize how differences in these phase diagrams arise from the dimensions of the rare regions. 

To build intuition, we consider success probabilities controlled by free energies in the context of maximum likelihood decoding. We neglect the random sign disorder in these stat mech models for simplicity, as we do not believe the sign disorder affects the relevant properties of the phases. At zero temperature (appropriate for describing the MWPM numerics in the main text), the phase transitions are in fact driven entirely by the random sign disorder; however, see the discussion in Appendix \ref{sec:app_MWPM_griffiths}.

We make use of Kramers-Wannier dualities where defect free energy costs are mapped to spin correlation functions in the dual Ising models. Within this formulation, we attribute the difference in the phase diagrams to whether the dimension of the rare regions are below or above the lower critical dimension of the dual Ising models. In particular, the fact that a two dimensional Ising model can order is responsible for the absence of a stretched exponential phase of the toric code in (2+1)D.



\subsection{Quasi-1D rare regions for repetition code in (1+1)D spacetime}\label{sec:app_2dKW}

As we explain in the main text, the stat mech model relevant to the repetition code in (1+1)D with non-uniform noise rates is the McCoy-Wu model (after neglecting weak random sign disorder). The $z$ direction corresponds to time, where faulty measurements are performed up to a time $t=L_z$, and perfect measurements (simulating readout via single-site measurements) are performed at $t=L_z$.

In the stat mech model, 
\begin{align}\label{eq:deltap}
    \psucc - \pfail = \frac{Z_{++}-Z_{+-}}{Z_{++}+Z_{+-}}
\end{align}
where $Z_{++}$ refers to fixing both the top and bottom boundary to be all $\sigma = 1$ and $Z_{+-}$ refers to fixing the top and bottom boundaries to have opposite orientation. Note that this quantity can be interpreted in two equivalent ways in this primal model. It is asymptotically $1$ whenever the free-energy cost of inserting a domain wall grows unboundedly in system size. It additionally can be rewritten as $\psucc - \pfail = \langle \sigma_{i,z=0} \sigma_{j,z=L_z} \rangle_\infty$ in a model without fixed boundary conditions but with infinite-strength horizontal couplings $K_x = \infty$ on the top and bottom boundaries. 

When thinking in terms of free-energy costs, it's useful to consider just
\begin{align}\label{eq:deltap}
    \pfail = \frac{Z_{+-}}{Z_{++}+Z_{+-}}
\end{align}
and in particular the ratio of $\frac{Z_{+-}}{Z_{++}} \leq 1$. Note that when $\frac{Z_{+-}}{Z_{++}}$ is small, $\pfail$ is $\sim \frac{Z_{+-}}{Z_{++}}$, and when $\frac{Z_{+-}}{Z_{++}} \sim 1$, $\pfail \sim \frac{1}{2}$.

\begin{figure}
    \centering
    \includegraphics[width = \linewidth]{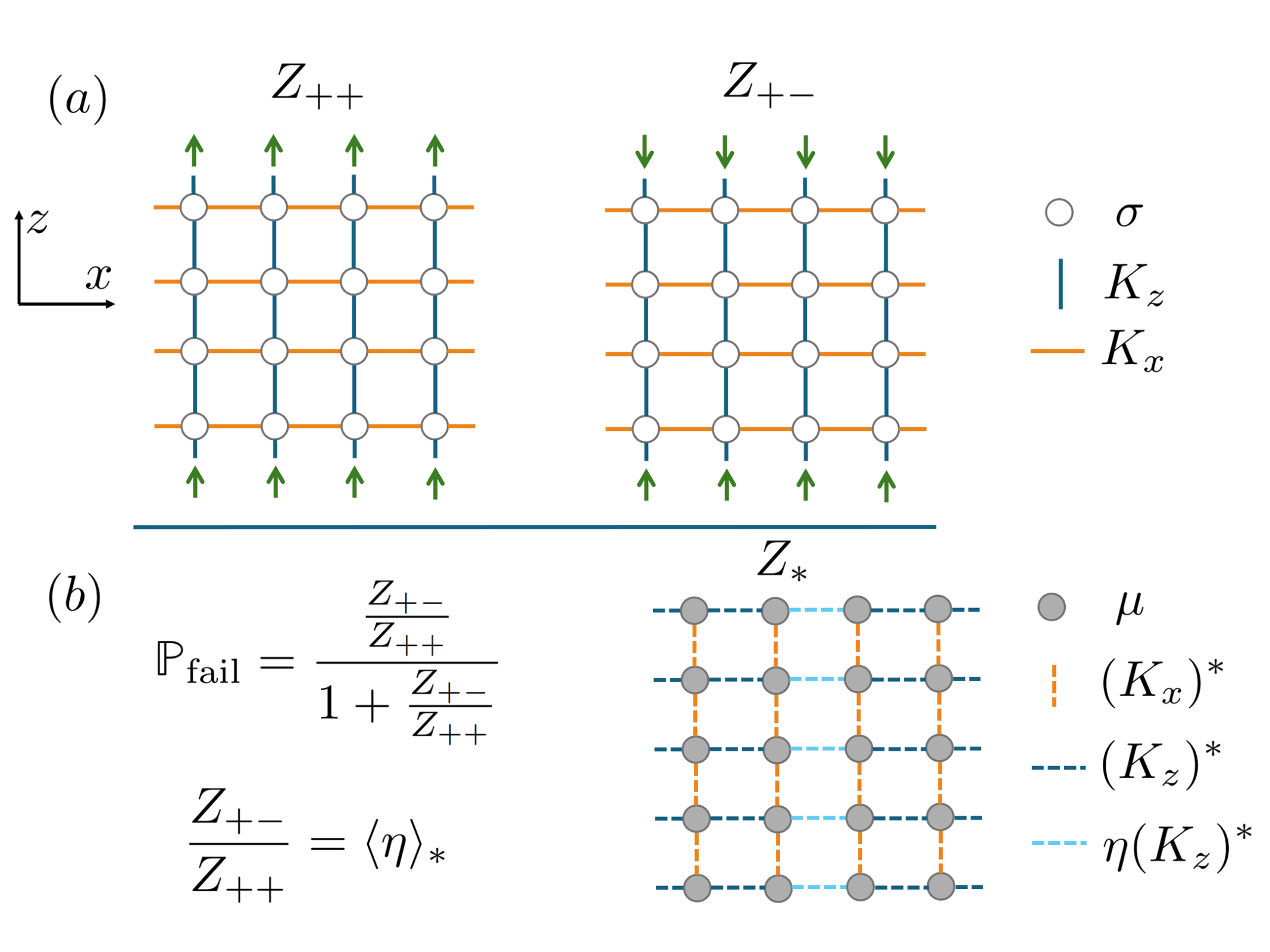}
    \caption{
    For the case of the repetition code in periodic boundary conditions, neglecting sign disorder:
    (a) depicts the models $Z_{++}$ and $Z_{+-}$ (see Eq.~\ref{eq:deltap} and surrounding discussion).  
    (b) shows what their ratio becomes under Kramers-Wannier duality, found by matching the high temperature expansion of the primal to the low temperature expansion of the dual. Dual spins $\mu$ live on the dual lattice, and the polarized boundary conditions at top and bottom of the primal model become open boundary conditions in the dual model. The duality requires an additional $\eta$ degree of freedom in the dual model that effectively toggles periodic and antiperiodic boundary conditions.}
    \label{fig:kwpbc}
\end{figure}

\begin{figure}
    \centering
    \includegraphics[width = \linewidth]{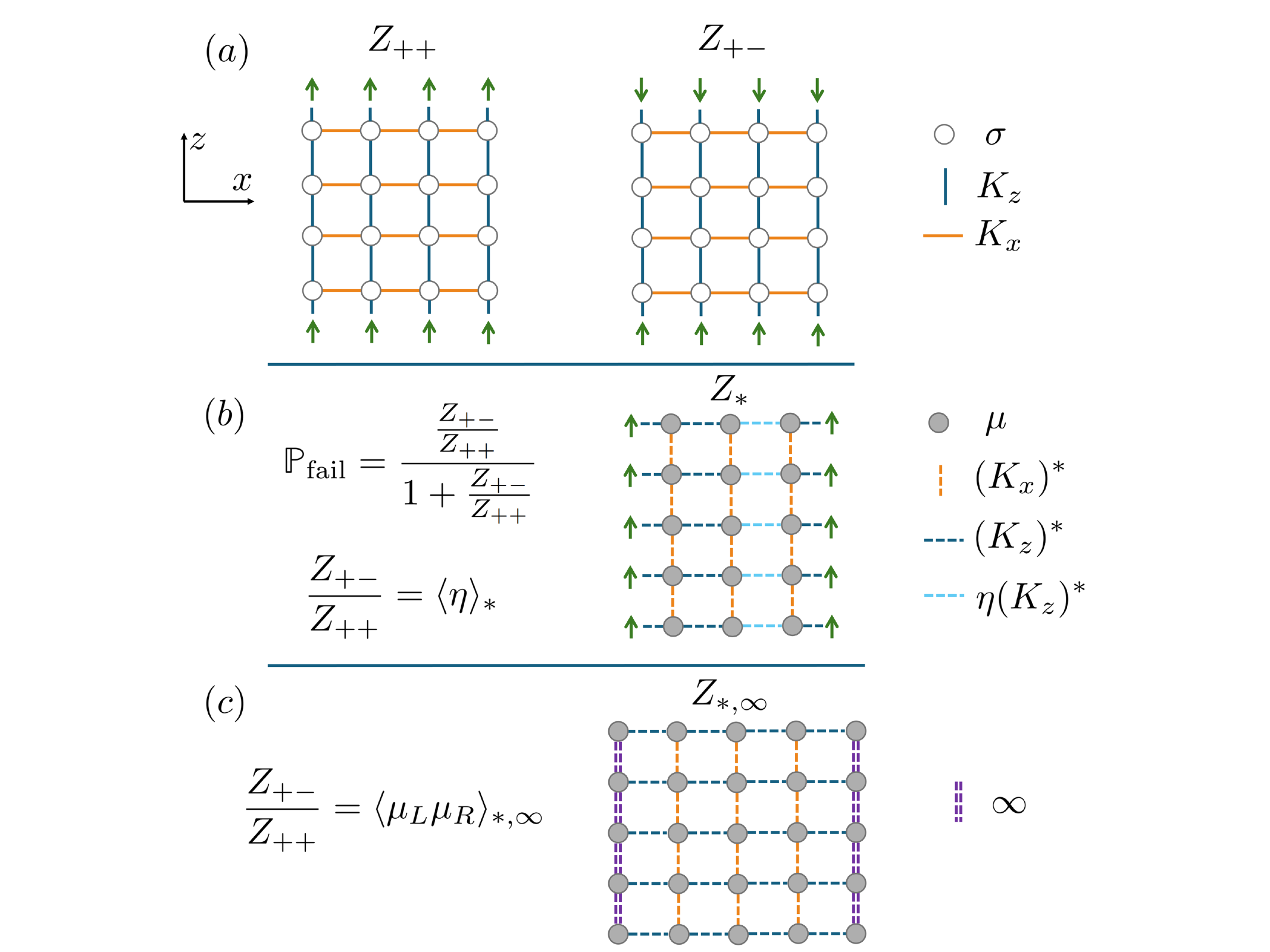}
    \caption{
    For the case of the repetition code in open boundary conditions, neglecting sign disorder:
    (a) depicts the models $Z_{++}$ and $Z_{+-}$ (see Eq.~\ref{eq:deltap} and surrounding discussion).  
    (b) shows what their ratio becomes under Kramers-Wannier duality, found by matching the high temperature expansion of the primal to the low temperature expansion of the dual. Dual spins $\mu$ live on the dual lattice, and the polarized boundary conditions at top and bottom of the primal model become open boundary conditions in the dual model. The open boundary conditions at left and right become polarized boundary conditions. The duality requires an additional $\eta$ degree of freedom in the dual model that effectively toggles periodic and antiperiodic boundary conditions.
    (c) gives an additional, equivalent representation in terms of correlation functions between dual spins on the left and right boundaries and does not require $\eta$. $\mu_L$ is any dual spin on the left boundary and $\mu_R$ is any dual spin on the right boundary. }
    \label{fig:kwobc}
\end{figure}

We can rewrite $\frac{Z_{+-}}{Z_{++}}$ in terms of quantities in the Kramers-Wannier dual models, and we detail this construction in Fig.~\ref{fig:kwpbc} and Fig.~\ref{fig:kwobc} for the respective cases of a periodic and open repetition code. For simplicity, in the figures we do not visually distinguish between rare and bulk regions; later, we will view these as uniform $L$ by $L_\perp$ systems that model the largest rare regions of height $L_\perp \sim \ln(L)$.

Under Kramers-Wannier duality, the spins of the dual model (which we will label as $\mu$ to distinguish them from the spins $\sigma$ of the primal model) live on the dual lattice, and local couplings $K = \beta J$ are in one-to-one correspondence with the dual couplings $K \to K^* = -\frac{1}{2}\log(\tanh(K))$.
The rare regions have $K_{x,\rm rare} < K_{c}^{2D} < K_{x,\rm bulk}$, see Fig.~\ref{fig:cartoons}(a).
Therefore, in the dual Ising model, the local couplings in the rare regions and in the bulk satisfy $K_{x, \rm rare}^\ast > K_{c}^{2D} > K_{x, \rm bulk}^\ast$. In the main text, $K_z$ is always set to $K_{c}^{2D}$ in all regions; more generally, $K_{x,c}^{2D}$ would then be a function of $K_z$.

For the repetition code in periodic boundary conditions, an additional $\eta$ ``domain wall" degree of freedom is required in the dual model. This $\eta$ variable allows the low temperature expansion of the dual model to match the high temperature expansion of the primal model; the latter includes lines that cross the bulk and terminate at the bottom and top boundaries. Such lines correspond to a \textit{single} vertical domain wall in the dual model. When the horizontal boundary conditions are periodic, domain walls naively come in pairs in the dual model. The introduction of $\eta$ allows for single domain walls in the dual model by converting a column of couplings in the dual model to be antiferromagnetic instead of ferromagnetic. In particular, as noted in Fig.~\ref{fig:kwpbc}, the couplings on such a column are weighted by $\eta$. Summing over $\eta=\pm 1$ thus sums over all configurations with an even and odd number of vertical domain walls in the dual model. 

In the high temperature expansion of $Z_{++}$, configurations with even and odd numbers of vertical lines crossing the bulk come with the same sign. However, such configurations come with opposite signs for $Z_{+-}$. By weighting the Boltzmann weights of the dual model by $\eta$, we reproduce the relative sign differences. 

That is,
\begin{align}
    \frac{Z_{+-}}{Z_{++}} = \langle \eta \rangle_* \,\,\,\,
\end{align}
where $*$ denotes an expectation in the dual model.

In our numerics, we always take periodic boundary conditions for the repetition code. However, the dual of the free energy cost of a domain wall can be massaged into a conceptually simpler form in open boundary conditions. We summarize the dual model for open boundary conditions in Fig.~\ref{fig:kwobc}. 

In Fig.~\ref{fig:kwobc}(b), we show the corresponding dual model. However, unlike periodic boundary conditions, this expression can be massaged into a more natural correlation function between dual spins on opposite edges.

By sending $\mu \to -\mu$ to the right or left of the $\eta$-weighted bonds, we can convert configurations with $\eta=-1$ into configurations with $\eta=1$ but oppositely polarized left and right boundary conditions. We can view these fully polarized boundary conditions as coming from free boundary conditions with infinite couplings. Furthermore, the relevant signs of configurations can be found by weighting with $\mu_L \mu_R$ instead of $\eta$; here $\mu_L$ is any dual spin on the left boundary and $\mu_R$ is any dual spin on the right boundary; the infinite couplings on the boundary make the choice immaterial.

Explicitly,
\begin{align}\label{eq:dualcorrfunc}
    \frac{Z_{+-}}{Z_{++}} = \langle \mu_{L} \mu_{\rm R} \rangle_{*,\infty} \,\,\,\, \text{ (OBC) }
\end{align}
for any choice of $\mu_{L}$ on the leftmost boundary and any choice of $\mu_{R}$ on the rightmost boundary. Here $\infty$ denotes that the left and right boundaries have infinite vertical couplings; see Fig.~\ref{fig:kwobc}(c).

From this relation, we can identify the line tension of the domain wall with the inverse correlation length of the dual Ising model, namely $\sigma = \left(\xi^{\ast}\right)^{-1}$.

Below, we focus on the largest rare regions (LRR). We assume that $\psucc - \pfail$ is controlled by configurations where the domain wall is entirely contained within largest paramagnetic rare region, where the domain wall gains the most entropy and where the line tension is the smallest. For conceptual simplicity, we use an $L$ by $L_\perp$ system as a model of the LRR. The horizontal couplings are all $(K_{x,\rm rare})^*$. 

The height of the largest rare region $L_\perp$ goes as $\ln(L)$, but it is worthwhile to consider $L_\perp$ fixed and finite first. Importantly, the dual Ising model is (quasi) one-dimensional and does not order at any nonzero temperature $(K_{x,\rm rare})^* < \infty$, and always has a finite correlation length $\xi^\ast$.
Correspondingly, the line tension in the primal model $\sigma = \left(\xi^{\ast}\right)^{-1}$ is nonzero at any non-infinite temeperature $K_{x,\rm rare} > 0$.

The infinite critical temperature for the vanishing of this domain wall line tension can be compared with the corresponding decoding problem at $L_\perp = O(1)$.
When introducing the random sign disorder back into the primal Ising model according to the Nishimori conditions, the primal model describes the decoding problem of a repetition code with length $L$ and run for $L_\perp$ time steps.
In this case, perfect state initialization and perfect final syndrome measurement are assumed, so that the domain wall can only fluctuate between $y \in [0, L_\perp]$.
It is easy to show that this code has threshold $p_{\rm th} = 0.5$ for any $L_\perp = O(1)$.

For both finite $L_\perp$ and $L_\perp = \ln(L)$, the $\mu$-$\mu$ correlation function can be obtained from a low temperature expansion within the dual model, yielding $(\xi^\ast)^{-1} \propto e^{-2K^\ast \cdot L_\perp}$.
This expansion is similar to that leading to Eq.~\eqref{eq:domain_wall_tension_expansion_appendixB}.
Alternatively, viewing the dual partition function as a path integral with the $x$ direction as a Euclidean time, the inverse correlation length is given by the splitting between ground state energies of the transfer matrix, i.e. a transverse field Ising model of length $L_\perp$.
This is again exponentially suppressed by $L_\perp$ when $K_{x, \rm rare}^\ast > K_{c}^{2D}$. 

This points to a difference between strictly finite $L_\perp$ and $L_\perp$ growing unboundedly with $L$; the latter can have an asymptotically vanishing inverse correlation length. For $L_\perp \sim \log(L)$ and $K_{x, \rm rare}^\ast > K_{c}^{2D}$, the inverse correlation length in the dual model vanishes as a power law in $L$. Correspondingly, the domain wall tension in the primal model decays algebraically $\frac{1}{L^z}$, and the domain wall cost goes as $L^{1-z}$, where $z$ is a function of the coupling strength. 

\subsection{Quasi-2D rare regions for toric code in (2+1)D spacetime}\label{sec:app_3dKW}

After neglecting weak random sign disorder, the toric code in (2+1)D spacetime is described by a $\mathbb{Z}_2$ lattice gauge theory in three dimensions, where the coupling strengths are uniform within each plane but may vary from plane to plane, see Fig.~\ref{fig:cartoons}.
We again focus on the LRR, and treat it as a quasi-2D system. We similarly assume that a homologically nontrivial flux loop along the $x$ or $y$ directions receives the most contribution from configurations where the flux loop is completely contained within the LRR. More precisely, we assume that the behavior of the phases can be understood in terms of the properties of the LRR, even if some of the phase boundaries change when considering the additional effects of smaller rare regions.

Under Kramers-Wannier duality, the LRR is described by a quasi-2D Ising model with dimensions $L_x \times L_y \times L_\perp$, where $L_x, L_y \to \infty$. The height of the largest rare region $L_\perp \sim \ln(L_z)$ will therefore slowly diverge with system size $L$, but it is again worthwhile to consider $L_\perp$ fixed and finite first.

Within the LRR we have $T_{\rm rare}^\ast < T_c^{3D, *}$, where $T_c^{3D, *}$ is the critical temperature of the dual 3D Ising model.
Under appropriate boundary conditions, we also have the relation Eq.~\eqref{eq:dualcorrfunc} and can identify $\sigma = (\xi^\ast)^{-1}$. The dual Ising model in the LRR can develop long range order at finite temperature, with a critical temperature denoted $T_c^{2D, \ast}(L_\perp)$.  $T_c^{2D, \ast}(L_\perp)$ increases monotonically with $L_\perp$, and approaches the $3$d transition temperature $T_c^{3D, *}$ as $L_\perp \to \infty$.
Thus, for any $T_{\rm rare}^\ast > T_c^{3D, *}$, we have $T_{\rm rare}^\ast > T_c^{2D, \ast}(L_\perp)$ for $L_\perp$ larger than a $T_{\rm rare}^\ast$-dependent constant. 
Correspondingly, whenever $T_{\rm rare}^\ast > T_c^{3D, *}$, the inverse correlation length and the domain wall line tension asymptotically vanish as $L_\perp$ increases.

In the toric code decoding problem with $L_x = L_y = L_z = L$, $L_\perp \propto \ln(L)$ and hence $L_\perp$ is unbounded above as $L$ grows. The above discussion about the approach of $T_c^{2D, \ast}(L_\perp)$ to $T_c^{3D, *}$ implies that whenever $\prare > p_c^{3D}$, logical errors will then proliferate and the code is not decodable.
The contrast with the repetition code case can  therefore be attributed to the dimension of the rare regions.

When $T_{\rm rare}^\ast < T_c^{2D, \ast}(L_\perp)$, the correlation length can either be obtained via an expansion along the lines of Appendix~\ref{sec:linetensions}, or via the ground state splitting of a transverse field Ising model of size $L_y \times L_z$, both yielding $ \ln \sigma \propto - L_y \cdot L_z$ for a defect in the $x$ direction.




\subsection{Transversal defects for toric code in (2+1)D spacetime }\label{sec:app_transversal}


In our time dependent error rate toric code, we obtained a gauge theory where the plaquette coupling magnitudes were completely correlated in the spatial direction. By duality, this is related to a 3D Ising model with bond coupling magnitudes which are correlated in planes. The 3D Ising model with planar defects is believed to have a ``smeared" phase transition~\cite{PhysRevB.69.174410}.
As each of the rare regions have infinite extent in two spatial dimensions, they are able to undergo phase transitions independently of the bulk. Then, as one lowers the temperature, different parts of the system order independently at different temperatures and the global order parameter develops smoothly from 0, when the smallest rare region orders. At the temperature where the smallest rare region orders, there is an essential singularity in the free energy. 

\begin{figure}
    \centering
    \includegraphics[width=\linewidth]{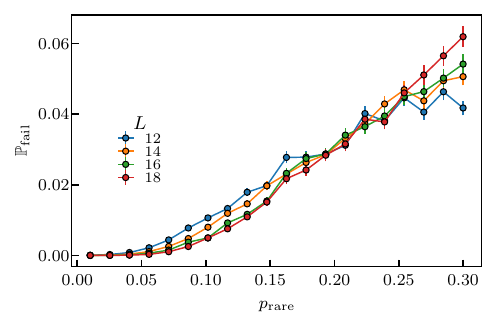}
    \caption{Failure rate due to transverse defect in a 2+1D toric code with planar disorder regions. Here we set the measurement error rate to $0.01$, $\gamma = 1/3$ and the average of $10^3$ error model realizations is shown. We used $10^4$ physical error configurations per point to determine $\pfail$. A crossing is observed near $\prare = 0.2$.}
    \label{fig:transverse}
\end{figure}

In Figure~\ref{fig:transverse}, we show numerics for the threshold of the transverse defect for the toric code in (2+1)D spacetime.
This defect goes in the temporal direction, perpendicular to the planar rare regions. In the repetition code, we were able to interpret the vanishing of the transverse defect as the point at which the underlying statistical mechanics model undergoes a phase transition. It is not clear whether that interpretation is true for the gauge theory. In future work, it would be interesting to further investigate this and whether symptoms of the smearing could be observed in the code. Furthermore, it would also be interesting to study what happens in the classical statistical mechanics model with planar defects when there is also uncorrelated sign disorder. 

\section{Crossover Scaling Function}\label{sec:app_crossover}

Here we discuss in some detail the crossover scaling function for an $L\times L \times L_\perp$ system with uniform error rate $p$.
We posit that the median failure probability of the code near $p_0^{3D}$ is captured by the following phenomenological crossover scaling function (see Eq.~\eqref{eq:scalingform1})
\begin{align} \label{eq:F1}
    \pfail(p, L, L_\perp) \approx \Phi\left[ (p-p_0^{3D}) \cdot L^{1/\nu_3}, L_\perp / L \right] \equiv \Phi[x,y].
\end{align}
As usual for finite size scaling, we require both $L$ and $L_\perp$ to be large for the scaling function to be descriptive.
We are particularly interested in the behavior of this function when $y \to 0$, as the largest rare regions will have typically $(L_\perp)_{\rm max} \propto \ln L$.

For future convenience, we define single-parameter scaling functions $\Phi^{3D}_y[z]$ for each $y$,
\begin{align} \label{eq:F2}
    \Phi^{3D}_y[z \equiv y^{1/\nu_3} \cdot x] \equiv \Phi[x,y].
\end{align}
They are therefore cross sections of $\Phi$ at constant values of $y$ (up to a rescaling of $x$).
For any finite $y > 0$, $\Phi^{3D}_y$ is \textit{analytic}, and describes the 3D bulk RPGT transition with aspect ratio $y$.

We extract the functional form of $\Phi^{3D}_{y \to 0}[z]$ in two steps.

(i)
We first take $L \to \infty$, while keeping $(p-p_0^{3D})$ and $L_\perp$ finite.\footnote{In this limit, we have $y \to 0$, $x \to \infty$, but $z = y^{1/\nu_3} \cdot x = (p-p_0^{3D}) \cdot L_\perp^{1/\nu_3}$ remains finite. This justifies our choice of the parameter $z$ for $\Phi_y^{3D}$, see Eq.~\eqref{eq:F2}.}
In this case, the system is an infinite 2D slab with height $L_\perp$, which has an $L_\perp$-dependent critical error rate, denoted $p_0^{2D}(L_\perp)$.
Therefore, as we take $L \to \infty$, we expect that
\begin{align} \label{eq:F3}
\pfail = 
\frac{3}{4} \cdot \Theta(p-p_0^{2D}(L_\perp)).
\end{align}
On the other hand, by definition of $\Phi^{3D}_y$ we have
\begin{align} \label{eq:F4}
    \pfail = \Phi^{3D}_{y=0}[z = (p-p_0^{3D}) \cdot L_\perp^{1/\nu_3}].
\end{align}
Comparing Eqs.~(\ref{eq:F3}, \ref{eq:F4}), we conclude that $\Phi^{3D}_{y=0}$ is necessarily \textit{singular}, with a step singularity at $z = z_0$.
Matching the location of the singularity, we have
\begin{align}
    & z_0 = (p_0^{2D}(L_\perp) - p_0^{3D}) \cdot L_\perp^{1/\nu_3} \\
    & \Rightarrow\quad p_0^{2D}(L_\perp) - p_0^{3D} = z_0 \cdot L_\perp^{-1/\nu_3},
\end{align}
see also Eq.~\eqref{eq:p0_Lperp}.
Again, this relation holds when $L_\perp$ is sufficiently large.
Therefore, from the crossover scaling function we can infer how $p_0^{2D}(L_\perp)$ approaches $p_0^{3D}$ with increasing $L_\perp$.

\begin{figure}[h]
    \centering
    \includegraphics[width=.8\linewidth]{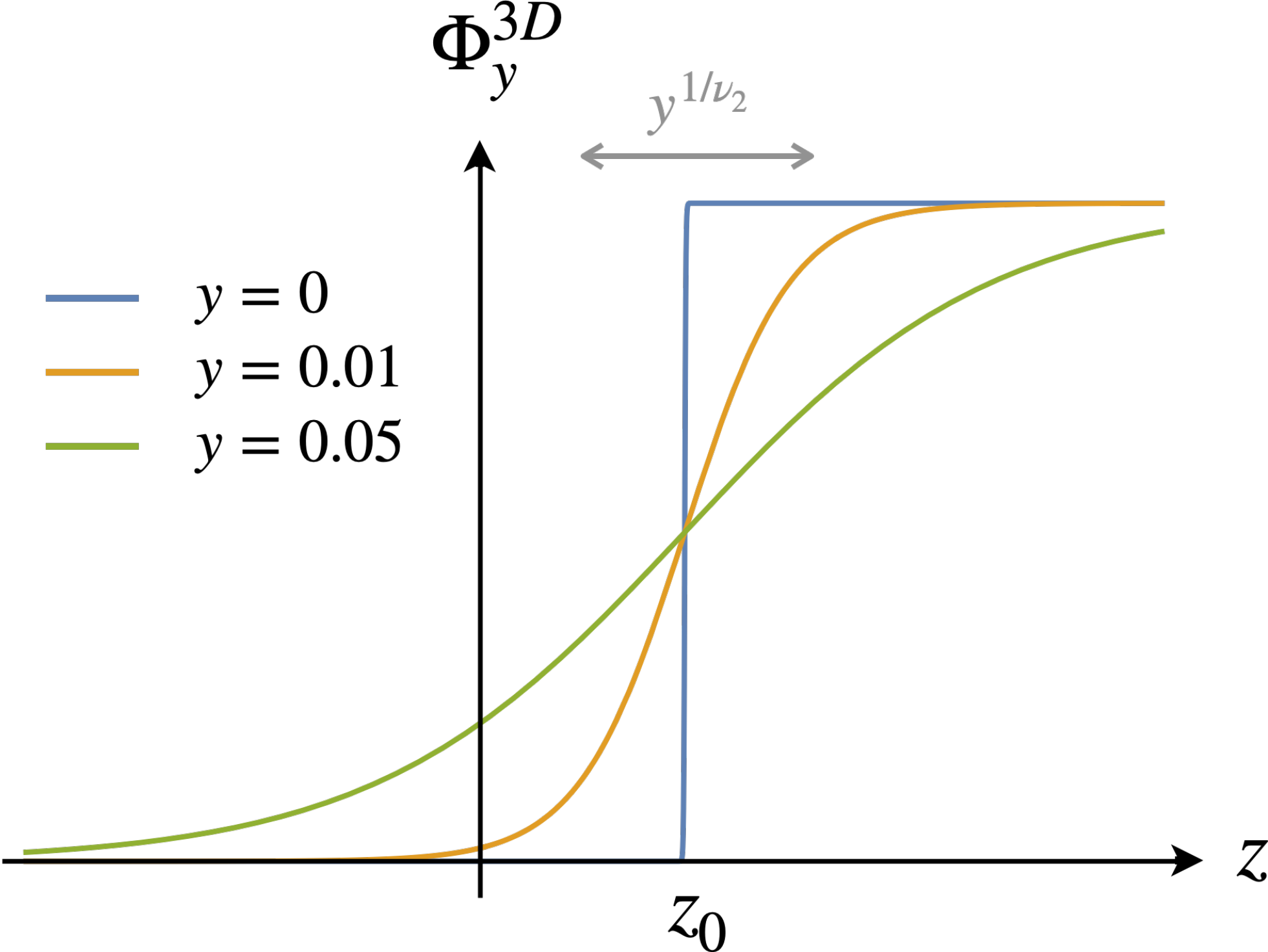}
    \caption{Schematic representation of the function $\Phi_y^{3D}[z]$, defined in Eq.~\eqref{eq:F2}.
    We are only plotting its functional form as $y \to 0$, as informed by Eq.~\eqref{eq:F12}.
    The function is analytic at any $y > 0$, as expected for a 3D bulk system with aspect ration $y$.
    At $y=0$, $\Phi_y^{3D}$ has a step singularity at $z = z_0$, as required by Eqs.~(\ref{eq:F3}, \ref{eq:F4}).
    Therefore, as $y \to 0$, the function ``sharpens up'' near $z = z_0$, with a width that shrinks as $y^{1/\nu_2}$.
    }
    \label{fig:Phi_y_3D}
\end{figure}

(ii)
Next, to see how $\Phi^{3D}_{y}$ becomes singular as $y \to 0$, we continue to keep $(p-p_0^{3D})$ and $L_\perp$ finite, and consider large but finite $L$, so that $y = L_\perp / L \ll 1$.
In this case, we expect to recover the scaling function near the 2D RBIM transition, namely
\begin{align}
    \pfail = \Phi^{3D}_{y \to 0}[z] = \Phi^{2D} [w = \lambda(L_\perp) \cdot (p-p_0^{2D}(L_\perp)) \cdot L^{1/\nu_2}],
\end{align}
where $\lambda(L_\perp)$ is an $L_\perp$-dependent multiplicative factor.
We emphasize that $\Phi^{2D}$ is an \textit{analytic} universal scaling function, and can be extracted e.g. from logical failure rates of the 2D toric code with perfect measurements.
Noticing that
\begin{align}
    & (p- p_0^{2D}(L_\perp))\cdot L^{1/\nu_2} \nn
    &= \left((p- p_0^{3D}) - (p_0^{2D}(L_\perp) - p_0^{3D} )\right) \cdot L^{1/\nu_2} \nn
    &= \left((p- p_0^{3D}) - z_0 \cdot L_\perp^{-1/\nu_3} \right) \cdot L^{1/\nu_2} \nn
    &= \left((p- p_0^{3D}) \cdot L_\perp^{1/\nu_3} - z_0\right) \cdot L_\perp^{-1/\nu_3} \cdot L^{1/\nu_2} \nn
    &= \left(z - z_0\right) \cdot y^{-1/\nu_2} \cdot L_\perp^{1/\nu_2-1/\nu_3},
\end{align}
we may therefore define as our scaling variable
\begin{align}
    w \equiv (z-z_0) \cdot y^{-1/\nu_2} = \lambda(L_\perp) \cdot (p- p_0^{2D}(L_\perp))\cdot L^{1/\nu_2}
\end{align}
where $\lambda(L_\perp) = L_\perp^{1/\nu_3-1/\nu_2}$,
and conclude that
\begin{align} \label{eq:F12}
&\pfail \stackrel{L_\perp / L \to 0}{=}
    \Phi^{3D}_{y \to 0}[z = y^{1/\nu_3} \cdot x] \nonumber\\
    &\quad\quad\quad = \Phi^{2D}[w = (z-z_0)\cdot y^{-1/\nu_2}],
\end{align}
see also Eq.~\eqref{eq:scalingform2}.
Based on Eq.~\eqref{eq:F12}, we include in Fig.~\ref{fig:Phi_y_3D} a schematic visualization of the function $\Phi^{3D}_y$ for small values of $y$ as $y \to 0$.

\subsection{Fluctuations in Size of Largest Rare Region}

In the main text we found that this scaling function Eq.~\eqref{eq:F12} collapsed the data for the planar disordered $L \times L \times L$ system, by setting $L_\perp \equiv \ln L$.
We claimed that this worked because the physics of the disordered model was entirely controlled by the largest rare region which was of height $(L_\perp)_{\rm max}\propto \ln L$. However, there are expected to be subleading $\mathcal{O}(1)$ fluctuations in the height of the largest rare region. It is worth then asking how these fluctuations affect the mean and median $\pfail$.

In particular, we argue that scaling collapse of the mean is asymptotically destroyed by these small fluctuations in $L_\perp$ as $L \to \infty$; however, this only noticeably occurs in systems with at least hundreds of millions of spins. On the other hand, the median is more resilient with scaling collapse maintained at all sizes. The only caveat is that the form of the scaling variable $w$ as a function of $L$ at similarly large system sizes will need to change slightly to include subleading-in-$L$ corrections to $L_\perp$.

For a fixed system size $L$, we may study the effect of a random $L_\perp$ through the scaling function Eq.~\eqref{eq:F12}, where the analytic scaling function $\Phi^{2D}$ now has a random parameter $w(L_\perp, L)$.
(For notational simplicity, we write $\Psi$ for $\Phi^{2D}$ henceforth.)
We expand $w$ around the mean of $L_\perp$ to the first order,
\begin{align}
    w(L_\perp, L) = w(\mathbb{E}[L_\perp], L) + \epsilon \cdot \frac{\partial w}{\partial L_\perp}\bigg|_{L_\perp = \mathbb{E}(L_\perp)} + O(\epsilon^2),
\end{align}
where $\epsilon \equiv L_\perp - \mathbb{E}[L_\perp]$.
We define $w_0 = w(\mathbb{E}[L_\perp], L)$, and write $w' \equiv (\partial w / \partial L_\perp)|_{L_\perp = \mathbb{E}(L_\perp)}$.
With these, we may write the mean failure probability as
\begin{align}
    & \mathbb{E}\left[ \Psi(w)\right] \nonumber \\
    &= \mathbb{E} \left[ \Psi(w_0 + \epsilon \cdot w') \right] \nonumber \\
    &=  \mathbb{E}\left[ \Psi(w_0) \right] + \mathbb{E}\left[ \epsilon \cdot w' \cdot \Psi'(w_0) \right] +\mathbb{E}\left[ \frac{\epsilon^2 \cdot (w')^2 }{2}\Psi''(w_0) \right] + \ldots \nonumber \\
    &= \Psi(w_0) +  \mathbb{E}[\epsilon] \cdot w'\cdot \Psi'(w_0) +\mathbb{E}\left[ \epsilon^2 \right] \cdot \frac{(w')^2}{2} \cdot \Psi''(w_0) + \ldots \nonumber \\
     &= \Psi(w_0) +\mathbb{E}\left[ \epsilon^2 \right] \frac{ (w')^2 }{2}\Psi''(w_0) + \ldots
\end{align} 
Here, we used $\mathbb{E}\left[ \epsilon \right]=0$, and we expect from standard extreme value statistics that $\mathbb{E}\left[ \epsilon^2 \right] = \mathcal{O}(1)$; by $\mathcal{O}(1)$, we mean that it is asymptotically independent of system size. 
We expect $\Psi''(w_0)$ to also be $\mathcal{O}(1)$, as $\Psi$ is analytic and bounded.

Note in particular that $\mathbb{E}\left[ \epsilon^2 \right] \frac{ (w')^2 }{2}\Psi''(w_0)$ (and the higher-order corrections) will generically depend on parameters like $L$ and error rates differently than $w_0$ depends on such parameters. This makes the expectation of the scaling function a sum of scaling functions with different scaling parameters, which will generically destroy single-parameter scaling if the corrections are not small.

Furthermore, we have that
\begin{align}
    w' = A L^{1 / \nu_2} L_\perp^{-1 / \nu_2 - 1} + B L^{1 / \nu_2} L_\perp^{-1/\nu_2 + 1/\nu_3 - 1}.
\end{align}
Here, $A$ and $B$ are constants which depend on $p - p_0^{3D}, z_0, \nu_2$ and $\nu_3$.
Notably, $w'$ diverges with $L$ even when the scaling argument $w$ is fixed and small.
Therefore, $\mathcal{O}(1)$ fluctuations in $L_\perp$ result in unbounded fluctuations in $w$, which will generically destroy the scaling collapse for the mean $\pfail$ when $L \to \infty$.

However, for $L_\perp = \ln(L)$, $w'<1$ for $L$ less than about $5000$, so this destruction of scaling collapse only occurs at quite large sizes. We still see scaling collapse of the mean of $\pfail$ at the sizes of $L$ that we probe (not pictured).

Because we do not expect single-parameter collapse for the mean in the thermodynamic limit, our plots of collapse are all for the median. We \textit{do} expect good single-parameter collapse of the median of $\pfail$ at all $L$.

In particular, for a monotonic function like $\Psi$, 
\begin{align}\label{eq:median}
    \mathrm{median}(\Psi(w(L,L_\perp))) = \Psi(\mathrm{median}(w(L,L_\perp)))
\end{align}
so collapse is maintained so long as the non-random $\mathrm{median}(w(L,L_\perp))$ is used as the scaling variable.

As a last technical note, the form of $\mathrm{median}(w(L,L_\perp))$ will slightly differ at sufficiently large sizes from what we use for collapse. We use $w = (p-p_c^{2d}(L_\perp)) L^{1/\nu_2} L_\perp^{1/\nu_3 - 1/\nu_2}\big|_{L_\perp = c \ln(L)}$, taking $L_\perp$ directly proportional to $\ln(L)$ and neglecting the subleading $\mathcal{O}(1)$ corrections. At large sizes $L$, the multiplicative factor of $L^{1/\nu_2}$ nevertheless makes the effect of these subleading corrections non-negligible; at sufficiently large $L$, single-parameter scaling collapse occurs for the slightly more complicated parameter $(p-p_c^{2d}(L_\perp)) L^{1/\nu_2} L_\perp^{1/\nu_3 - 1/\nu_2}\big|_{L_\perp = c \ln(L) + \mathcal{O}(1)}$ with explicit $\mathcal{O}(1)$ corrections included in $L_\perp$. However, this minor distinction only matters once the spins number in the hundreds of millions.


\bibliography{main}

\end{document}